\documentclass[fleqn,usenatbib]{mnras}
\usepackage{amsmath,amssymb,natbib,latexsym,times}

\usepackage[T1]{fontenc}
\usepackage{aecompl}

\usepackage{lineno}

\usepackage{verbatim}
\usepackage{hhline}
\usepackage[usenames,dvipsnames]{xcolor}
\usepackage{todonotes}
\usepackage{xspace}
\usepackage{hyperref}

\usepackage{eso-pic}
\AddToShipoutPictureBG*{%
  \AtPageUpperLeft{%
    \hspace{0.75\paperwidth}%
    \raisebox{-3.5\baselineskip}{%
      \makebox[0pt][l]{\textnormal{DES 2019-0493}}
}}}%

\AddToShipoutPictureBG*{%
  \AtPageUpperLeft{%
    \hspace{0.75\paperwidth}%
    \raisebox{-4.5\baselineskip}{%
      \makebox[0pt][l]{\textnormal{Fermilab PUB-20-254-AE}}
}}}%

\newcommand\edit[1]{#1}

\numberwithin{equation}{section}

\DeclareMathSymbol{:}{\mathord}{operators}{"3A}

\newcommand{\epsf}{\ensuremath{e_\textsc{psf}}}
\newcommand{\Tpsf}{\ensuremath{T_\textsc{psf}}}

\newcommand{\bfx}{\ensuremath{\mathbf{x}}}
\newcommand{\bfxpt}{\ensuremath{\mathbf{x} + \boldsymbol{\theta}}}

\newcommand{\piff}{\textsc{Piff}}
\newcommand{\ngmix}{\textsc{ngmix}}
\newcommand{\pixmappy}{\textsc{Pixmappy}}
\newcommand{\sex}{\textsc{SExtractor}}
\newcommand{\psfex}{\textsc{PSFEx}}
\newcommand{\sklearn}{\textsc{scikit-learn}}
\newcommand{\treegp}{\textsc{TreeGP}}
\newcommand{\galsim}{\textsc{GalSim}}

\newcommand*\justify{%
  \fontdimen2\font=0.4em
  \fontdimen3\font=0.2em
  \fontdimen4\font=0.1em
  \fontdimen7\font=0.1em
  \hyphenchar\font=`\-
}

\newcommand\code[1]{\texttt{\small\justify #1}}

\usepackage{listings}
\usepackage[dvipsnames]{xcolor}

\newcommand\YAMLcommentstyle{\scriptsize \color{blue}\mdseries}
\newcommand\YAMLcolonstyle{\scriptsize \color{purple}\bfseries}
\newcommand\YAMLkeystyle{\scriptsize \color{black}\bfseries}
\newcommand\YAMLvaluestyle{\scriptsize \color{ForestGreen}\bfseries}
\newcommand\ProcessThreeDashes{\llap{\color{black}\mdseries-{-}-}}
\makeatletter
\newcommand\language@yaml{yaml}
\expandafter\expandafter\expandafter\lstdefinelanguage
\expandafter{\language@yaml}
{
  keywords={true,false,null,y,n},
  keywordstyle=\color{darkgray}\bfseries,
  basicstyle=\YAMLkeystyle,                                 
  sensitive=false,
  comment=[l]{\#},
  morecomment=[s]{/*}{*/},
  commentstyle=\YAMLcommentstyle,
  stringstyle=\YAMLvaluestyle,
  moredelim=[l][\color{orange}]{\&},
  moredelim=**[il][\YAMLcolonstyle{:}\YAMLvaluestyle]{:},   
  moredelim=**[il][\ \YAMLcolonstyle{\{}\YAMLkeystyle]{\{\ },   
  moredelim=**[il][\YAMLcolonstyle{\}}\YAMLkeystyle]{\}},
  moredelim=**[il][\ \YAMLcolonstyle{[}\YAMLkeystyle]{[\ },
  moredelim=**[il][\YAMLcolonstyle{]}\YAMLkeystyle]{]},
  moredelim=**[il][\YAMLcolonstyle{,}\YAMLkeystyle]{,},
  morestring=[b]',
  morestring=[b]",
  literate =    {---}{{\ProcessThreeDashes}}3
                {>}{{\textcolor{red}\textgreater}}1     
                {|}{{\textcolor{red}\textbar}}1 
                {\ -\ }{{\bfseries\ -\ }}3,
}
\lst@AddToHook{EveryLine}{\ifx\lst@language\language@yaml\YAMLkeystyle\fi}
\makeatother

\newcommand\eqn[1]{equation~\ref{#1}}

\newcommand\eqnc[2]{equations~\ref{#1}~--~\ref{#2}}

\newcommand\fig[1]{Figure~\ref{#1}}

\newcommand\app[1]{Appendix~\ref{#1}}

\newcommand\sect[1]{\S \ref{#1}}

\allowdisplaybreaks

\title[DES Year 3 Results: PSF Modeling]{Dark Energy Survey Year 3 Results:\\Point-Spread Function Modeling}

\author[DES Collaboration]{
\parbox{\textwidth}{
\Large
M.~Jarvis,$^{1}$\thanks{mjarvis@physics.upenn.edu}
G.~M.~Bernstein,$^{1}$
A.~Amon,$^{2}$
C.~Davis,$^{2}$
P.~F.~L\'eget,$^{3}$
K.~Bechtol,$^{4}$
I.~Harrison,$^{5}$
M.~Gatti,$^{6}$
A.~Roodman,$^{2,7}$
C.~Chang,$^{8,9}$
R.~Chen,$^{10}$
A.~Choi,$^{11}$
S.~Desai,$^{12}$
A.~Drlica-Wagner,$^{8,13,9}$
D.~Gruen,$^{14,2,7}$
R.~A.~Gruendl,$^{15,16}$
A.~Hernandez,$^{14}$
N.~MacCrann,$^{11,17}$
J.~Meyers,$^{18}$
A.~Navarro-Alsina,$^{19,20}$
S.~Pandey,$^{1}$
A.~A.~Plazas,$^{21}$
L.~F.~Secco,$^{1}$
E.~Sheldon,$^{22}$
M.~A.~Troxel,$^{10}$
S.~Vorperian,$^{23}$
K.~Wei,$^{9}$
J.~Zuntz,$^{24}$
T.~M.~C.~Abbott,$^{25}$
M.~Aguena,$^{26,20}$
S.~Allam,$^{13}$
S.~Avila,$^{27}$
S.~Bhargava,$^{28}$
S.~L.~Bridle,$^{5}$
D.~Brooks,$^{29}$
A.~Carnero~Rosell,$^{30,31}$
M.~Carrasco~Kind,$^{15,16}$
J.~Carretero,$^{6}$
M.~Costanzi,$^{32,33}$
L.~N.~da Costa,$^{20,34}$
J.~De~Vicente,$^{35}$
H.~T.~Diehl,$^{13}$
P.~Doel,$^{29}$
S.~Everett,$^{36}$
B.~Flaugher,$^{13}$
P.~Fosalba,$^{37,38}$
J.~Frieman,$^{13,9}$
J.~Garc\'ia-Bellido,$^{27}$
E.~Gaztanaga,$^{37,38}$
D.~W.~Gerdes,$^{39,40}$
G.~Gutierrez,$^{13}$
S.~R.~Hinton,$^{41}$
D.~L.~Hollowood,$^{36}$
K.~Honscheid,$^{11,17}$
D.~J.~James,$^{42}$
S.~Kent,$^{13,9}$
K.~Kuehn,$^{43,44}$
N.~Kuropatkin,$^{13}$
O.~Lahav,$^{29}$
M.~A.~G.~Maia,$^{20,34}$
M.~March,$^{1}$
J.~L.~Marshall,$^{45}$
P.~Melchior,$^{21}$
F.~Menanteau,$^{15,16}$
R.~Miquel,$^{46,6}$
R.~L.~C.~Ogando,$^{20,34}$
F.~Paz-Chinch\'{o}n,$^{47,16}$
E.~S.~Rykoff,$^{2,7}$
E.~Sanchez,$^{35}$
V.~Scarpine,$^{13}$
M.~Schubnell,$^{40}$
S.~Serrano,$^{37,38}$
I.~Sevilla-Noarbe,$^{35}$
M.~Smith,$^{48}$
E.~Suchyta,$^{49}$
M.~E.~C.~Swanson,$^{16}$
G.~Tarle,$^{40}$
T.~N.~Varga,$^{50,51}$
A.~R.~Walker,$^{25}$
W.~Wester,$^{13}$
and R.D.~Wilkinson$^{28}$
\begin{center} (DES Collaboration) \end{center}
}
\vspace{0.4cm}
\\
\parbox{\textwidth}{
The authors' affiliations are shown in Appendix \ref{sec:affiliations}.
}
}

\date{Accepted XXX. Received YYY; in original form ZZZ}

\pubyear{2020}

\begin{document}

\label{firstpage}
\pagerange{\pageref{firstpage}--\pageref{lastpage}}
\maketitle
\begin{abstract}

We introduce a new software package for modeling the point-spread function (PSF) of astronomical images,
called \piff\ (PSFs In the Full FOV), which we apply to the first three years (known as Y3) of the
Dark Energy Survey (DES) data.
We describe the relevant details about the algorithms used by \piff\ to model the PSF, including how the
PSF model varies across the field of view (FOV).
Diagnostic results show that the systematic errors from the PSF
modeling are very small over the range of scales that are important for the DES Y3 weak lensing analysis.
In particular, the systematic errors from the PSF modeling are significantly smaller than the corresponding
results from the DES year one (Y1) analysis.
We also briefly describe some planned improvements to \piff\ that we
expect to further reduce the modeling errors in future analyses.

\end{abstract}


\section{Introduction}
\label{sec:intro}

The Dark Energy Survey \citep[DES]{des2016} has already produced very precise constraints on cosmology
\citep{y1-keypaper} using just the first year of data (Y1).
The Y1 weak lensing (WL) cosmic shear measurements alone were able to constrain the
combination $\sigma_8 \left(\Omega_m/0.3\right)^{0.5}$ to 3.5\% uncertainty \citep{y1-shearcorr}.
Such precise constraints require that systematic uncertainties be controlled to levels smaller
than the statistical uncertainties. The first three years of DES data (collectively referred to as Y3) thus require even
better control of the various systematic effects that impact shear measurements.

One of the most significant systematic uncertainties in the DES Y1 cosmic shear analysis was the
estimation of the point-spread function (PSF) at the location of each galaxy.  PSF estimation
is difficult because the PSF varies both spatially across the field of view (FOV) and
temporally from one exposure to the next.  
The FOV of the Dark Energy Camera \citep[DECam;][]{DECam} is quite large,
covering a diameter of 2.2 degrees and containing 62 2k$\times$2k CCDs (charge coupled devises).
These CCDs are not perfectly coplanar, which means that the PSF variation
is moderately discontinuous at the edges of each CCD.
Furthermore, the PSF can be directly
measured only at the locations of stars, which are essentially point sources, so their
surface brightness profiles are direct measurements of the PSF at those locations.
Since the galaxies are observed at different locations, the PSF must be interpolated
to the locations of the galaxies.  

In the Y1 analysis effort, it was clear that while the PSF estimation was sufficiently accurate
to not significantly bias the Y1 cosmic shear results, we would need to make some improvements
for the Y3 analysis, which has smaller statistical uncertainties.  Other recent cosmic shear
experiments have similarly montioned PSF modeling errors as a significant source of
systematic uncertainty \citep{Hamana20, Hildebrandt20}.

The Y1 PSF estimation used the software package \psfex\ \citep{BertinPSFEx2011}.
This package estimates the PSF at the location of each star using a linear combination of
basis vectors.  In our case, the basis used was simply the pixel values on a grid.
The coefficients of this model are interpolated using a polynomial in the chip
coordinates, $x$ and $y$, across
the area of each CCD image.

This method worked very well and is quite robust.  However,
\citet*{y1-shearcat} detected some aspects that could potentially be improved for future DES analyses.
First, the mean measured size of the PSF models showed a non-negligible offset from the 
mean measured size of the input stars.
Second, a map of both size and shape residuals vs location on the focal plane
exhibited spatial patterns that were clearly related to features of the astrometric distortion solutions
for DECam.

To address both of these issues, we developed new PSF estimation software, named
\piff\ (PSFs In the Full FOV), which we describe in this paper. The
principal goals for how to extend beyond the capabilities of the
tried-and-true \psfex\ software were:
\begin{itemize}
  \item Description of the PSF in sky coordinates rather than pixel
    coordinates to better account for the effects of astrometric distortion.
   \item Potential to fit the PSF simultaneously across multiple
      detectors in the FOV in cases where continuous behavior across
      detectors could be exploited for better fitting.
    \item Ability to use astrometric maps that are not expressible in
      the FITS (Flexible Image Transport System) standards\footnote{
      \url{https://fits.gsfc.nasa.gov/fits_wcs.html}}.
    \item Code written in or easily accessible from Python.
    \item Modular design to enable new (potentially non-linear)
      PSF models, outlier rejection algorithms, and interpolation
      schemes.
\end{itemize}

The version of \piff\ used for the DES Y3 WL analysis is 0.2.4, but \piff\ has been in continuous development 
since the Y3 PSF models were finalized. 
In this paper, we demonstrate diagnostics for the DES Y3 PSF model based on version 0.2.4, 
but we will also point out various features that have already been improved upon in later versions.
Throughout the paper, we will mention the choices used for the DES Y3 WL
analysis where we discuss the various options enabled by \piff.  The complete
configuration file that was used for this analysis is given in \app{config}.

We give an overview of the procedure for building a PSF model for an
exposure in \sect{sec:scheme}.
We describe the options
in \piff\ for the PSF surface brightness model, interpolation schemes, and how it finds the
overall solution in \sect{sec:model}--\ref{sec:psf}.  We describe the DES Y3 data in \sect{sec:data} and
show our tests of the PSF solution on these data in \sect{sec:tests}.  In \sect{sec:future},
we discuss some potential future improvements to \piff\ currently in development, and
we conclude in \sect{sec:conclusion}. Detailed instructions for using
\piff\ can be found in the online code documentation\footnote{\url{http://rmjarvis.github.io/Piff/}}.

\section{Scheme of operation}
\label{sec:scheme}

At any particular location in the field of view, the point-spread function is a two-dimensional function
describing the mapping from a delta function (a point) in the sky at
coordinates $(u_0,v_0)$ to a surface brightness profile
measured by the detector at coordinates $(x_0,y_0)$ as $I_\mathrm{image}(x-x_0,y-y_0).$
The functional form of this mapping is called the ``model'' in \piff.

In \piff, we usually express the model in sky coordinates, for which we use the
notation $(u,v)$, rather than
image (pixel) coordinates, for which we use $(x,y)$. The features of the PSF
created by atmospheric and optical distortions tend to vary more smoothly
across the field of view when considered in sky coordinates than 
in pixel coordinates.  This is especially true in the presence of
high-frequency components of astrometric distortion such as 
``tree rings'' \citep{Kotov10, plazas14} which originate in the
detector.  While detector-based components of the PSF (such as charge
diffusion) may or may not have this property, the DES modeling has been
found to be
overall better-behaved when using sky coordinates.  Fitting in pixel
coordinates remains an option in \piff\ for applications where this is not true. 
A mixed model is left to future development. 

The sky coordinates $u$ and $v$ are defined in the local tangent
plane projection of the sky, as seen from Earth, around a nominal position of the source.
Positive $v$ is to the north, and 
positive $u$ is west.  Converting the surface brightness profile
between image and sky coordinates
is straightforward given the knowledge of the world coordinate system (WCS),
which defines the functions $u(x,y)$ and $v(x,y)$.
\begin{linenomath*}
\begin{equation}
I_\mathrm{image}(x,y) = I_\mathrm{sky}(u,v) 
\left| \begin{matrix} 
\frac{du}{dx} & \frac{dv}{dx} \\
\frac{du}{dy} & \frac{dv}{dy}
\end{matrix} \right|
\end{equation}
\end{linenomath*}
where the last factor is the determinant of the Jacobian of the coordinate transformation,
which we identify as the pixel area, $A_\mathrm{pix}$.

We normalize this function to have unit flux\footnote{
This seems an obvious choice, but one should be careful when using the \piff\ PSF
model for photometry, in terms of how this normalization interacts with the zero point calibration of
aperture fluxes},
\begin{linenomath*}
\begin{equation}
  \int du\,dv\,I_\mathrm{sky}(u,v) = \int dx\,dy\,I_\mathrm{image}(x,y) = 1.
  \label{psfnorm}
\end{equation}
\end{linenomath*}
Henceforth, we will dispense with the ``sky'' label and merely
use $I(u,v)$ as the surface brightness profile in sky coordinates.

The function $I(u,v)$ here is taken to include a convolution by the pixel response.
This is sometimes referred to as the ``effective PSF'' \citep[ePSF;][]{Bernstein02},
\edit{which we take to be continuous even though it is only sampled at the pixel centers}.
It is an implementation detail of the various models in \piff\ whether the underlying
description is natively the ePSF or the PSF profile without the pixel.
Different models handle this differently, but all models know how to properly
draw themselves onto an image including the correct application of the pixel response.

The data for star $i$ consist of the (sky-subtracted) counts $\hat
d_{i\alpha}$ in all pixels indexed by $\alpha$ in a region around star
$i$.  If the pixels each subtend an area $A_\mathrm{pix}$ on the sky,
then the 
model for the observed surface brightness of star $i$ with flux $f_i$ and sky position
$(u_i,v_i)$ is
\begin{linenomath*}
\begin{equation}
  d_{i\alpha} = f_i A_\mathrm{pix} I(u_{i\alpha}{-}u_i,v_{i\alpha}{-}v_i).
\label{dmodel}
\end{equation}
\end{linenomath*}
The likelihood of obtaining the data given the model is given by
\begin{linenomath*}
\begin{equation}
  -2 \log \mathcal{L}(\hat{\mathbf d}_i) = \sum_{\alpha \in i}
                                           \left[ \frac{\left(\hat d_{i\alpha} {-}
                                           d_{i\alpha}\right)^2}{
                                           \sigma_{i\alpha}^2 {+}
                                           d_{i\alpha}} + \log\left( \sigma_{i\alpha}^2 {+}
                                           d_{i\alpha}\right) \right],
\label{likelihood}
\end{equation}
\end{linenomath*}
where we have assumed that the model and data are in units of
photoelectrons, so that the total variance of a pixel is the sum of
the read/background variance $\sigma^2_{i\alpha}$ and the Poisson variance
from the expected counts, $d_{i\alpha}$.

The PSF model will always be a function of some vector of parameters
$\mathbf{p}$.  We will denote as $\mathbf{\hat p}_i$ the parameters
selected to fit the data for star $i$.  When we refer to individual
parameters, we will use $k$ to index them: $\hat p_{ik}$. 

We must also define some
interpolation scheme to provide a function $\mathbf{p}(u,v,c)$,
which can return $\mathbf{p}$ at an
arbitrary location $(u,v)$ in the sky domain of the exposure and
potentially depend on additional properties of the target, which we
denote as $c$ since this is most likely to be a color.  The
interpolated function will be somehow trained on the particular fits
$\mathbf{\hat p}_i$ measured at the $(u_i, v_i, c_i)$ of the PSF
stars. 

One basic sequence of operations for a \piff\ run for a given exposure
(or portion thereof) is:
\begin{enumerate}
  \item Read as input the list of stellar sky coordinates
    $(u_i,v_i)$ for candidate PSF-fitting stars $i=1,\ldots,N_\star,$
    pixel data $\hat{\mathbf d}_i$ for the regions around each star,
    and noise levels $\sigma_{i\alpha}$ at each pixel.  We may also have
    other potential data on the stars upon which the PSF can depend,
    in particular a color $c_i.$
  \item Specify a parametric form $I(u,v |\mathbf{p})$ for the PSF model.
  \item Specify a method to interpolate the PSF parameters $p_k$ from the
    positions (and perhaps colors) of the stars to an arbitrary test point $(u,v,c)$ in
    the domain.  \label{step:singlestar}  
  \item Execute a fit of the model to the pixel
    data at each star to yield a maximum-likelihood estimate
   $\mathbf{\hat p}_i$ at the location of the star. \label{step:interp}
   \item Train or fit the interpolation with the $\mathbf{\hat p}_i$ vectors.
   \item Possibly identify and excise outlier stars that are deemed to be
    poor exemplars of the PSF according to some metric.
    \item Iterate from step~\ref{step:singlestar} to refit and reject
      outliers until convergence is reached.
\end{enumerate}

An alternative approach, described in
\sect{sec:basis} and also implemented in \psfex, is to merge steps
\ref{step:singlestar} and \ref{step:interp} into a direct solution for
the parameters of the interpolating function, bypassing the single-star
solutions. This has a higher level of computational complexity, but
enables the use of PSF models that are not fully specified by a single
star, e.g if the model has significant power below the Nyquist scale for the given pixel size,
or if any stars have missing data.

We will detail the components of the procedure in the following
sections.  First, though, we note that
there is a subtlety to the process: there is a
degeneracy between the choice of center $(u_i,v_i)$ for a star and
the functional form of $I(u,v)$.  We can always shift the nominal center,
insert a countering shift into the model, and obtain the same fit.
There are two possible means to break this degeneracy.  \piff\ allows
one to choose whichever is more appropriate, based on the nature of the
input centers.

The first mode, which we label ``centered-PSF,'' is to
force the model to be centered, 
i.e. $\int du\,dv\, \{u,v\} I(u,v)=0$, and treat the positions $(u_i,v_i)$
of the stars as free parameters.  This is appropriate when the
stellar positions and/or the WCS solution are not trusted to be extremely accurate, so
offsets are more likely indicative of errors in the input rather than real PSF centroid motion.
In this mode, the centroid parameters of the PSF model are forced to zero
and are not included in the interpolation.  It would be left to other
processing steps to evaluate any variable astrometric displacements
across the exposure.  This option is the default in \piff, or it can be
explicitly specified by setting \code{centered=True} when defining a \code{Model}.
This is the mode that we used for the DES Y3 WL analysis.

The second mode, which we will call ``fixed-star,'' is to trust the input positions $(u_i,v_i)$ of the
stars, and assign the shift freedom to the parameters
of the PSF function $I(u,v)$.  This would be
appropriate if the input positions and WCS are both known to be
extremely accurate.
In this case the centroid offset can be taken to be due to stochastic
atmospheric refraction or other instrumental effects, which
can then be interpolated along with the other PSF parameters.
This option is enabled by setting \code{centered=False} in \piff\ when
defining a \code{Model}.

\section{PSF Model}
\label{sec:model}

\piff\ provides a number of possible choices for the functional form of the model, $I(u,v)$,
with different advantages and disadvantages in terms of
simplicity and realism.

\subsection{Analytic radial profiles}
\label{sec:analytic}

The simplest PSF models in \piff\ are based on isotropic radial functions of the intensity.
This radial function may be sheared and dilated, but is otherwise fixed to a given
functional form:.
\begin{linenomath*}
\begin{align}
  I_p(u,v) &= f(r), \\
  r &= \sqrt{{u^\prime}^2 + {v^\prime}^2}, \\
  \left( \begin{matrix} u^\prime \\ v^\prime \end{matrix} \right) &=
   \frac{s}{\sqrt{1-g_1^2-g_2^2}}
                    \left( \begin{matrix} 1+g_1 & g_2 \\
                                             g_2 & 1-g_1 \end{matrix} \right)
  \left( \begin{matrix} u \\ v \end{matrix} \right)
\label{eq:fr_uv}
\end{align}
\end{linenomath*}
where $g_1$, $g_2$, and $s$ are free parameters in the fit, which effect the shear
and dilation\footnote{
We only have three degrees of freedom in the $2 \times 2$ matrix,
since any rotation component is irrelevant for an initially isotropic function.}.

Normally, these analytic profiles are taken to describe the PSF profile without the
pixel response, so the full ePSF profile, which we've been calling $I(u,v)$, would be
\begin{align}
I(u,v) = I_p(u,v) \ast P(u,v)
\end{align}
where $\ast$ denotes convolution and $P(u,v)$ is the square pixel response projected
into sky coordinates according to the WCS.  This convolution is handled automatically
by \galsim\footnote{\url{https://github.com/GalSim-developers/GalSim}} \citep{galsim2015}
when the profile is being drawn.

However, there is also an option to tell \galsim\ not to include the pixel convolution when drawing
the PSF by setting \code{include\_pixel=False}, in which case $I(u,v) = I_p(u,v)$.  
This is not normally recommended (not least because the
pixel response is not a radial function), but it can be useful in some very simple scenarios
for testing purposes.

There are three available options in \piff\ for the radial function, $f(r)$:
\begin{itemize}
\item
\code{Gaussian} uses the radial function
\begin{linenomath*}
\begin{equation}
f_\mathrm{Gaussian}(r) = A \exp{\left(-\frac{r^2}{2 \sigma^2}\right)},
\end{equation}
\end{linenomath*}
where $\sigma$ is taken to be 1 arcsecond, since it is degenerate with the overall dilation that is allowed by the fit.
Gaussian profiles are not usually particularly good matches for real PSFs, but they are very simple,
and therefore are sometimes useful for simulations.

\item
\code{Moffat} uses the radial function
\begin{linenomath*}
\begin{equation}
f_\mathrm{Moffat}(r) = A \left(1 + (r/r_0)^2\right)^{-\beta},
\end{equation}
\end{linenomath*}
where $r_0$ is degenerate with the dilation so is essentially arbitrary at this point\footnote{
The choice of $r_0$ in \piff\ is such that the half-light radius is 1 arcsecond.}.
The $\beta$ parameter controls the concentration of the profile and must be explicitly specified.
We do not yet have the capability to fit for $\beta$ as part of the fit, although this could be added
if someone has a use case that requires it.

\item
\code{Kolmogorov} is defined in Fourier space \citep{Racine96}
\begin{linenomath*}
\begin{equation}
F_\mathrm{Kolmogorov}(k) = A \exp{\left(-\left(\frac{24 \Gamma(6/5)}{5} \right)^{5/6}  \left(\frac{\lambda k}{2\pi r_0} \right)^{5/3}\right) },
\end{equation}
\end{linenomath*}
where $r_0$ here is the Fried parameter \citep{Fried66}, a property of the atmospheric conditions at the time of the
observation. 
The real-space function is the Hankel transform of $F_\mathrm{Kolmogorov}(k)$:
\begin{linenomath*}
\begin{equation}
f_\mathrm{Kolmogorov}(r) = \frac{1}{2\pi} \int_0^\infty F_\mathrm{Kolmogorov}(k) J_0(k r) k dk.
\end{equation}
\end{linenomath*}

\end{itemize}

In all cases, the overall amplitude $A$ is set such that the integrated
flux is unity.  Because the radial functions are naturally centered,
implementing the centered-PSF mode is simple.  To induce fixed-star mode,
we need to add two additional parameters $(u_c,v_c)$ to the set of model parameters to
specify how the center is offset from the nominal position.  These are subtracted from $(u,v)$
prior to the application of the shear and dilation (\eqn{eq:fr_uv}).

The maximum likelihood solution is found for 6 parameters:
\begin{linenomath*}
\begin{equation}
\mathbf{\hat p_\mathrm{full}} = \left[ A, g_1, g_2, s, u_c, v_c \right] 
\end{equation}
\end{linenomath*}
using non-linear least squares (with \code{scipy}).  The estimated covariance
matrix of the solution is computed from the Jacobian matrix $\mathbf{J}$ returned by
\code{scipy}:
\begin{linenomath*}
\begin{align}
\mathrm{Cov}\left(\mathbf{\hat p_\mathrm{full}}\right) &= \left(\mathbf{J}^T \mathbf{J}\right)^{-1}
\label{gsmodel_cov}
\end{align}
\end{linenomath*}
which may be used by the interpolator if needed (currently only \code{GPInterp}
can use it; cf. \sect{sec:gp}).

Then the final parameter vector for the PSF model is a subset of the full maximum likelihood
solution:
\begin{linenomath*}
\begin{equation}
\mathbf{\hat p} = 
 \begin{cases}
   ~~ \left[ g_1, g_2, s \right] & \quad \textrm{centered-PSF} \\
   ~~ \left[ g_1, g_2, s, u_c, v_c \right] & \quad \textrm{fixed-star}
 \end{cases}
\end{equation}
\end{linenomath*}
where the flux of the star $A$ is ignored.  
For centered-PSF mode, the fitted $u_c, v_c$ values are used to update the
nominal position of the star.  

\subsection{Pixel grid}
\label{sec:pixelgrid}

The \code{PixelGrid} model is much more general than the analytic profiles.  It models the PSF profile
as a two-dimensional grid of points, smoothed into a
continuous function by a 1d kernel function $K(x):$
\begin{linenomath*}
\begin{equation}
  I(u,v) = \sum_{k=1}^{N_\mathrm{pix}} p_k K(u{-}u_k) K(v{-}v_k).
\end{equation}
\end{linenomath*}
There are $N_\mathrm{pix}$ free parameters $p_k$ in the model, where $N_\mathrm{pix}$
is the total number of pixels in the model grid.  The center ($u_k$,
$v_k$) of each pixel $k$ are set onto a regular square grid of chosen
spacing and dimension.  The grid orientation is parallel to the $u$ and $v$ directions.

For the kernel function, we use the Lanczos interpolation kernels, $K(x) = L_n(x)$, where
\begin{linenomath*}
\begin{align}
L_n(x) &\equiv 
\begin{cases} 
1 & \mathrm{if} ~|x| = 0\\
\frac{n}{\pi^2 x^2} \sin \left(\pi x\right) \sin \left(\frac{\pi x}{n}\right) & \mathrm{if} ~0<|x|<n \\
0 & \mathrm{if} ~|x| \ge n
\end{cases}
\label{eq:lanczos}
\end{align}
\end{linenomath*}
and $n$ is a free (integer) parameter, whose default in \piff\ is $n=3$.

The grid of pixels where the model is defined does not need to be the same size or orientation as the pixels in the
observed image.  Indeed, since the model is constructed in sky coordinates, there is always some WCS
transformation from model space to the data, which means that in general an interpolation would always be
required to the constrain the model parameters defined in $(u,v)$ coordinates from data in $(x,y)$ coordinates.

This is the model we used for the DES Y3 WL analysis.  Specifically, we used a $17 \times 17$ grid of
pixels with a pixel scale of 0.30 arcseconds.  These are about 15\% larger than the data pixels (0.263 arcsec),
which we found helped improve the robustness of the fit compared to using model pixels nearly the same
size as the data pixels.

The coefficients $p_k$ for a given star can be constrained by minimizing
\begin{linenomath*}
\begin{align}
\chi^2 &= \sum_{\alpha} 
\frac{\left(d_\alpha - f A_\mathrm{pix} I(u_\alpha{-}u_c,v_\alpha{-}v_c) \right)^2}
{\sigma_\alpha^2 + d_\alpha}
\label{eq:chisq_pixelgrid}
\end{align}
\end{linenomath*}
where the sum on $\alpha$ is over the observed data pixels and the star has some flux $f$ and
centroid $(u_c,v_c)$. 
Minimizing this leads straightforwardly to a design matrix for the coefficients $\{p_k\}$
\begin{linenomath*}
\begin{align}
\mathbf{A} \mathbf{p} &= \mathbf{b}
\end{align}
\end{linenomath*}
for which the maximum likelihood solution is
\begin{linenomath*}
\begin{align}
\mathbf{\hat p} &= \left(\mathbf{A}^T \mathbf{A}\right)^{-1} \mathbf{A}^T \mathbf{b} \\
\mathrm{Cov}\left(\mathbf{\hat p}\right) &= \left(\mathbf{A}^T \mathbf{A}\right)^{-1}
\end{align}
\end{linenomath*}

Since the model has translational freedom, the centered-PSF
mode described in \sect{sec:scheme} is 
not as simple as it was for radial profiles.
The centroid
of the pixel grid profile is a derived property based on all $N_\mathrm{pix}$
parameters---as is the overall flux constraint.
To implement this mode, the current version of \piff\ starts with an
initial estimate of the flux and centroid based on simple (0th and 1st order) moments
\edit{of the data, $\hat d_{i\alpha}$}.
The model centroid is not forced to zero during the fit, and the flux is not forced
to unity, so the solution can have non-zero centroid and non-unit flux.
Then, at the start of each iteration, the position of each star is updated such that 
the best match to the current model would have zero centroid.  
Similarly, the model is renormalized to have unit flux, and the flux of the star is updated to match
the best fit to this model.
This is essentially a projected gradient descent algorithm on these three parameters.
This algorithm tends to converge quickly to models with zero centroid and unit flux in almost all cases.
The fixed-star mode uses the same pattern,
but only updating the flux, which tends to converge even faster.

\begin{figure}
\includegraphics[width=\columnwidth]{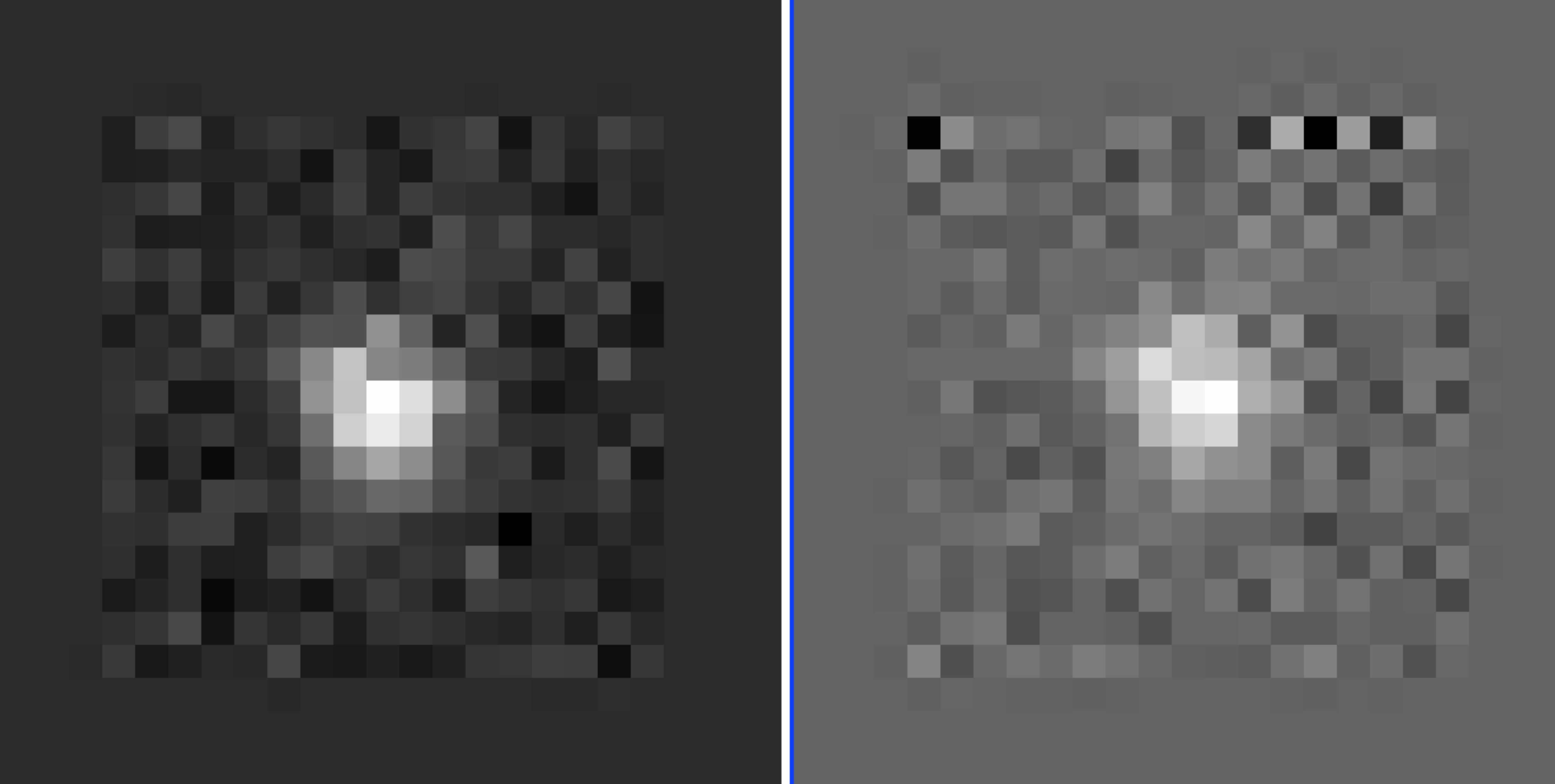}
\caption{
An example of a ``checkerboard'' failure mode that can sometimes happen with PixelGrid solutions
with version 0.2.4 of \piff.
On the right is the model for the PSF in one of the Y3 DES images,
extrapolated to a location far from
any of the constraining stars.  Notice the checkerboard pattern, especially near 
near the upper right corner of the model.  On the left is the PSF for the same location of the same image using
\piff\ version 0.3.0.  The checkerboard pattern is no longer present.
\label{fig:checkerboard}
}
\end{figure}

The version of \piff\ that we used for the Y3 analysis (version 0.2.4) used a
different algorithm for the centered-PSF mode.  The details of this
algorithm are given in \app{constrainedcentroid}.
However, we found that this could sometimes lead to
numerical instabilities in the solutions during the interpolation
step, leading to spurious checkerboard patterns when extrapolated 
to locations not near any constraining stars (cf. \fig{fig:checkerboard}).
This failure mode was mitigated by using a
model grid size that was somewhat larger than the data pixel size (0.3 arcseconds
vs 0.263 arcseconds), which made the checkerboard failures rare.
We believe the blacklist
procedure described in \sect{sec:blacklist} removed most of the CCDs 
that were still affected by it.  The algorithm described above does not
suffer from this failure mode, but we did not discover this solution until after
the Y3 PSF solutions were finalized.
\edit{We expect that future applications of \piff\
will be able to use a grid size commensurate or even smaller than the data pixels,
although this has yet to be extensively tested.}

\section{PSF Interpolation}
\label{sec:interp}

The profile of the PSF is not constant across the field of view.  We have measurements of the PSF
at the locations of the stars, but we generally need to know the PSF at other locations, such as where
galaxies are observed.  

\piff\ has a number of potential methods for doing the interpolation, which can typically be matched
with any of the various PSF models described in \sect{sec:model}.  The information about the
model is parameterized by a vector $\mathbf{p}$ with measurements (or constraints) taken at the 
locations of the stars $\mathbf{\hat p}_i(u_i,v_i)$.

All currently-implemented interpolation schemes except the
\code{BasisPolynomial} (\sect{sec:basis}) follow the separated-solution
  scheme of \sect{sec:scheme}, whereby a maximum-likelihood solution
  $\mathbf{\hat p}_i$ is first derived for each star for the specified model.
Then a second solution is derived for the parameters of the
interpolated function ${\mathbf p}(u,v,c)$ by fitting to or training
on the $\mathbf{\hat p}_i$ vectors.


\subsection{Simple polynomial interpolation}
\label{sec:poly}

The simplest possible interpolation scheme is to take the model as constant across the field of view.
This is not usually a particularly good approximation to reality, but it can
be useful in some cases, such as a simulation that really does have a constant PSF model.
This simplistic interpolation scheme is called \code{Mean} in \piff.

Slightly more complicated, but still rather simple, is \code{Polynomial} interpolation.  In this scheme,
each coefficient in the vector $\mathbf{p}$, $p_k$, is interpolated according to an arbitrary
polynomial in $u$ and $v$ (and possibly other parameters assigned to
each star, such as color).
\begin{linenomath*}
\begin{equation}
p_{ik} = \sum_{m} Q_{km} K_{im}(u_i, v_i, \dots)
\end{equation}
\end{linenomath*}
where $\mathbf{K}_{i}$ is a basis vector giving the relevant polynomial terms
at the location of star $i$: ${\mathbf K}_i=\{1, u_i, v_i, u_i^2, u_i v_i, v_i^2, \dots\}$.
Note that the basis may include other factors, such as ones involving a color term $c_i$, if desired.  
Terms in the $u$ and $v$ parameters are included up to a 
maximum total $m{+}n$.
The order of the polynomial may be different for
each parameter if desired.  The coefficients $Q_{km}$ are found by a
maximum-likelihood fit to the $\hat p_{ik}$ values.

\begin{figure*}
\includegraphics[width=0.24\textwidth]{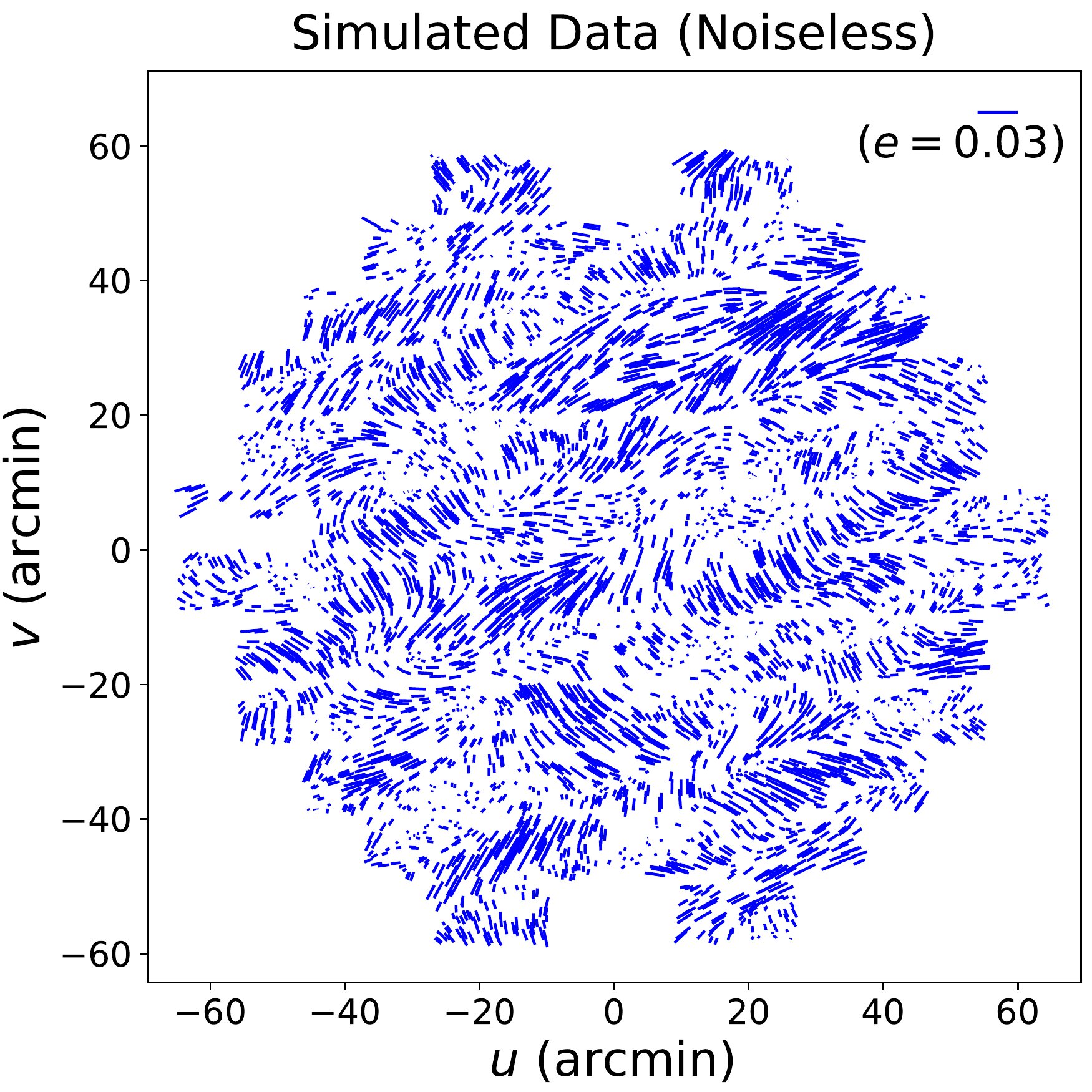}
\includegraphics[width=0.24\textwidth]{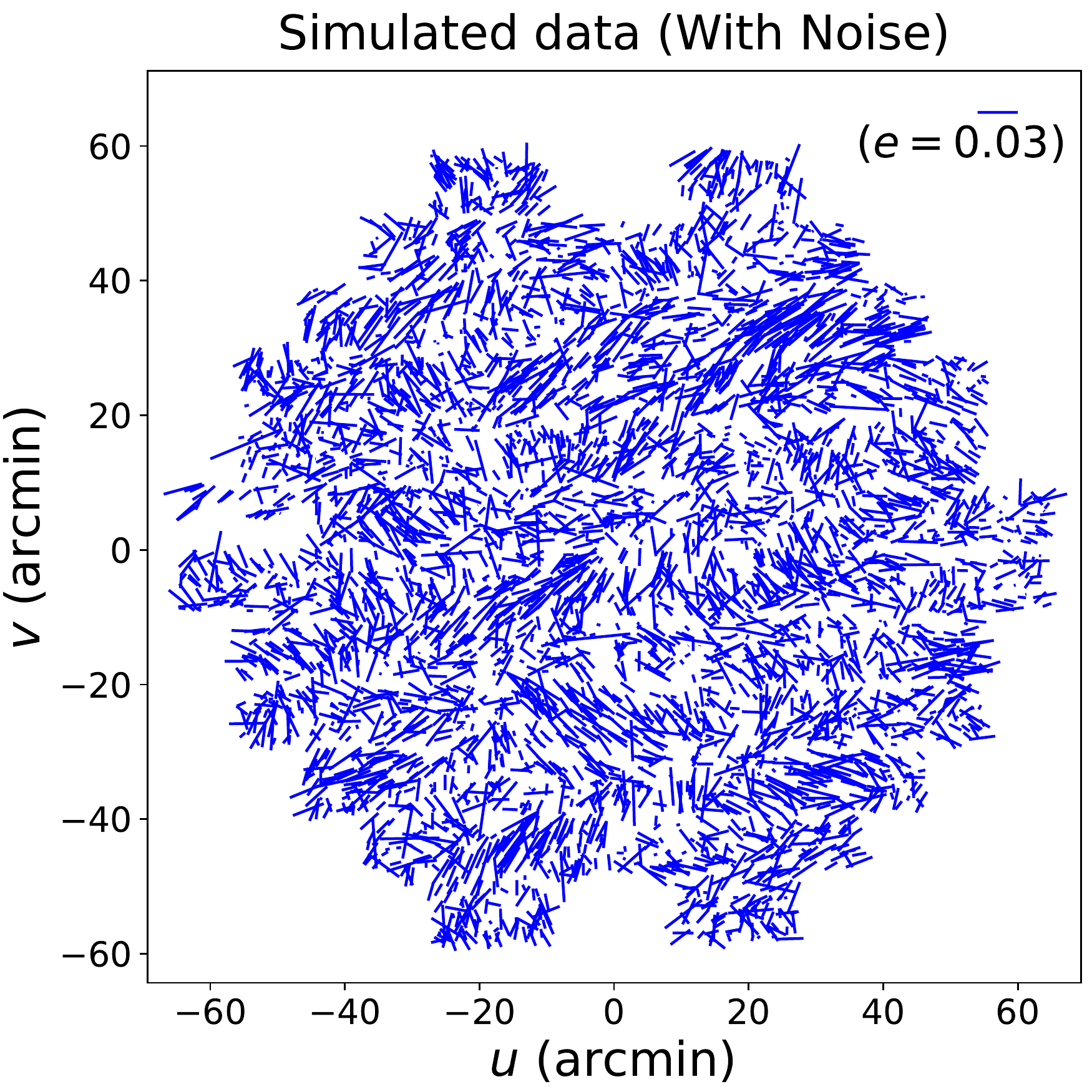}
\includegraphics[width=0.24\textwidth]{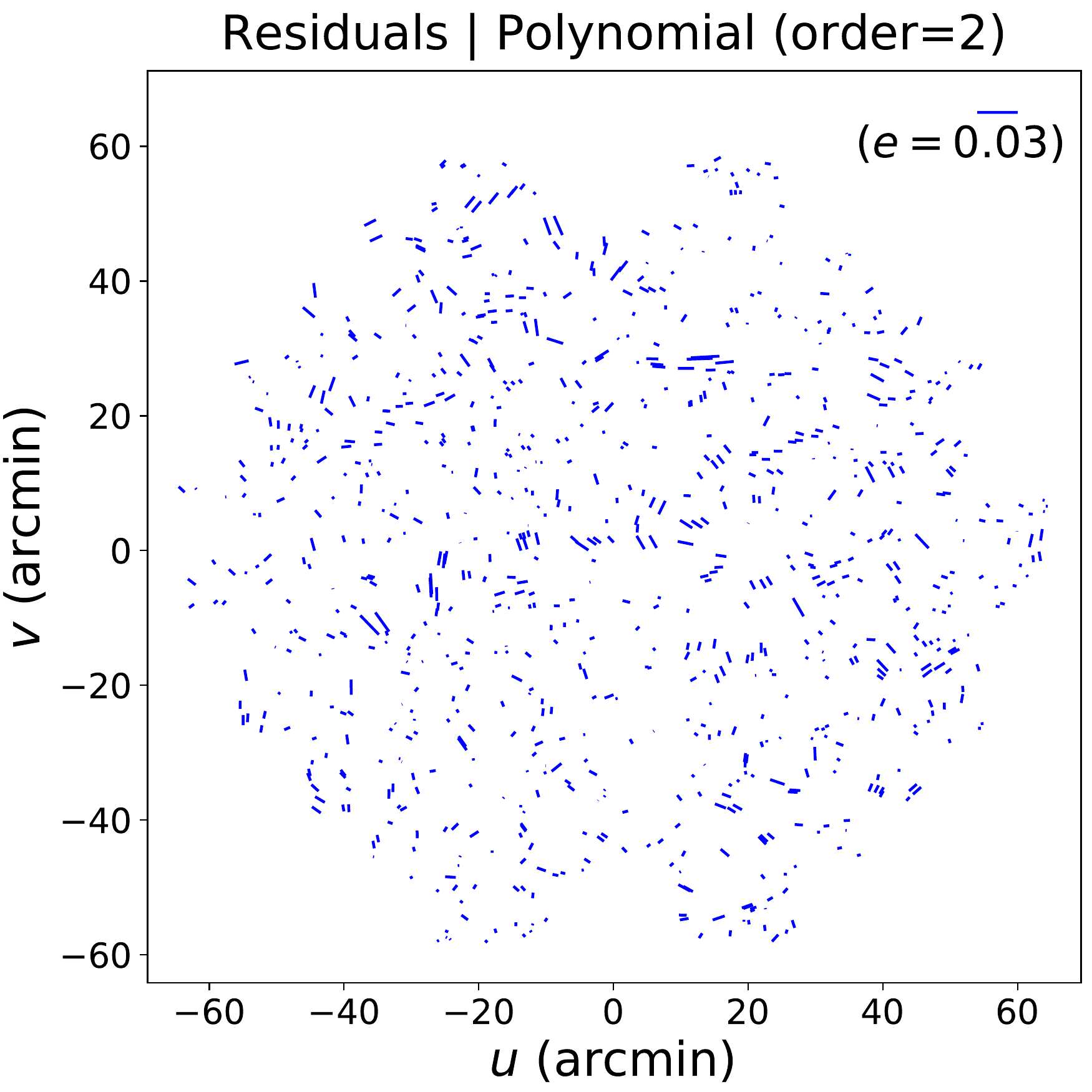}
\includegraphics[width=0.24\textwidth]{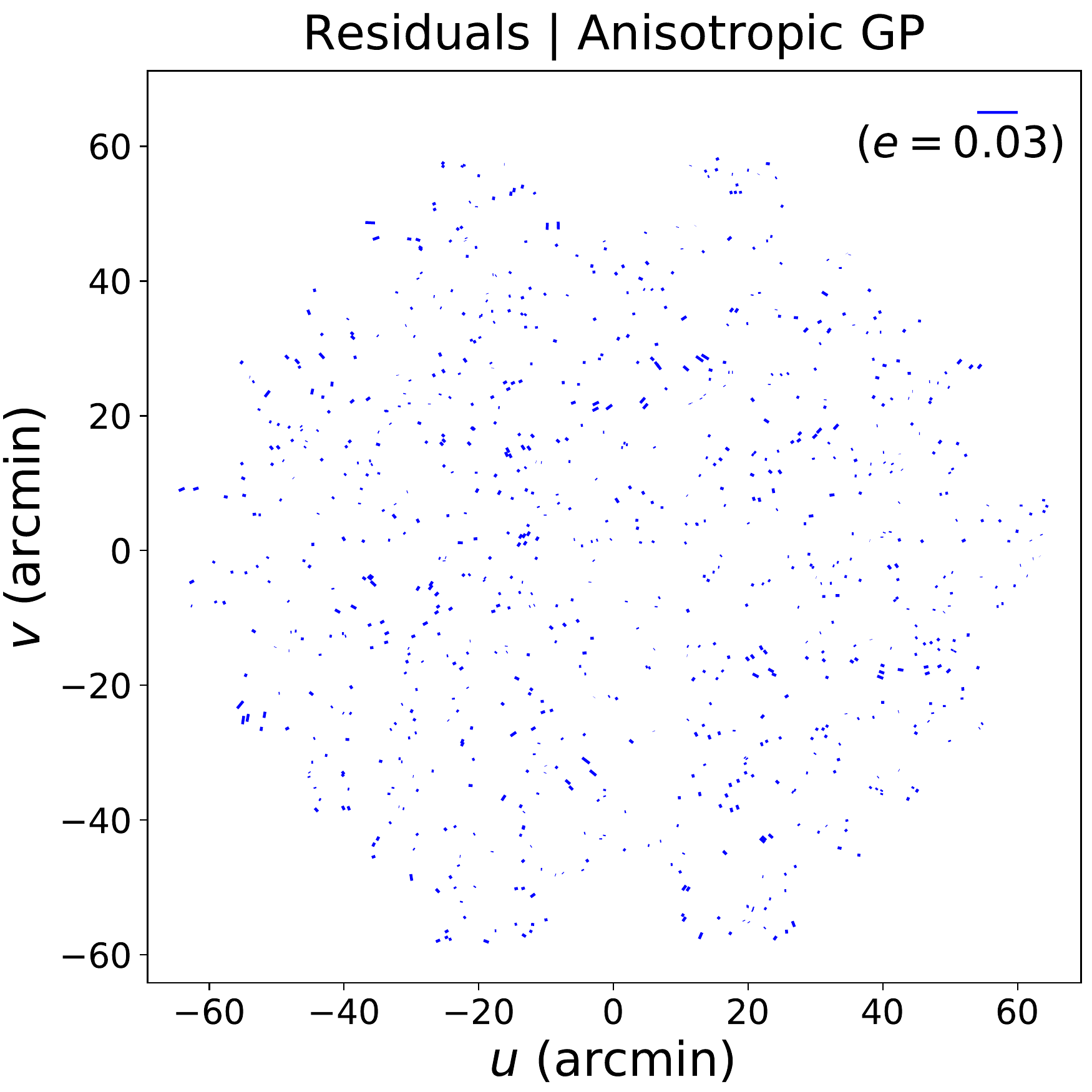}
\caption{
A simulation of a purely atmospheric PSF pattern.  From left to right the panels show 
(1) the true PSF shapes without noise, 
(2) the shapes with measurement noise, 
(3) the residuals from a second order polynomial fit on each CCD, 
and (4) the residuals from an anisotropic Gaussian process fit over the whole FOV.  
Each panel shows ``whiskers'' whose length is proportional to the ellipticity of the PSF,
and whose orientation matches the orientation of longest axis of the PSF shape.
In both cases, the residuals are shown only for a set of reserve stars, which were not
used to fit the PSF model.
\label{fig:gp}
}
\end{figure*}

\subsection{K-nearest neighbors}
\label{sec:knn}

The \code{kNNInterp} class in \piff\ implements a k-nearest neighbor regression \citep{KNN} on the
parameter vectors $\mathbf{\hat p}_i$.  The estimated parameter vector at an arbitrary location
$(u,v)$ is taken to be the weighted average of the $k$ constraining stars nearest to that location.

The implementation is based on the \sklearn\footnote{
\url{https://scikit-learn.org/}} class \code{KNeighborsRegressor},
which can weight the $k$ points either uniformly or inversely by their distance to the
interpolation location.  The default weight is uniform, which we find usually gives better
results for PSF interpolation, since it smooths over typical noise in the measurements
of the PSF vectors.

This interpolation scheme can \edit{potentially} give better performance for PSF patterns that have
complicated functional forms, which are not well modeled by a polynomial.
The appropriate value of $k$ to choose (default is 15) depends on both the stellar
density of the observations and the expected scale length of variations in the
true PSF pattern.

\subsection{Gaussian processes}
\label{sec:gp}

The \code{GPInterp} class in \piff\ implements a Gaussian process regression \citep{GP} on the
parameter vectors $\mathbf{\hat p}_i$.  Gaussian process regression, 
also known as ``Kriging,'' 
assumes that the parameter vector $\mathbf{p}$ at every location
is drawn from a multi-dimensional Gaussian random field across the $(u,v)$ space.
It requires an estimate of the spatial covariance function
of $\mathbf{p}$, commonly referred to as the kernel.
The interpolation estimate at an arbitrary location $(u,v)$ is the
minimum variance unbiased estimate from the Gaussian distribution at that location conditioned
on the values $\mathbf{\hat p}_i$ measured at all the PSF stars. 

The \code{GPInterp} class is implemented using the \treegp\ module\footnote{
\url{https://github.com/PFLeget/treegp}}.
It can use a number of possible kernels from \sklearn\
to define the covariance matrix along with a few custom options, which we use in \piff.
The default kernel is the so-called "squared exponential" or "radial basis function"
kernel, known as \code{RBF} in \sklearn.

The \treegp\ module has several methods for how to optimize the kernel's
hyper-parameters, which are available in \piff\ via the \code{optimizer} parameter:
\begin{enumerate}
\item
\code{likelihood} is the traditional maximum likelihood optimization.
It finds the hyper-parameters that maximize the likelihood of the Gaussian
process solution.  This is similar to the optimization done by the
\code{GaussianProcessRegressor} in \sklearn,
although technically \treegp\ has a custom implementation,
which uses \code{scipy.optimize} for its back end.
\item
\code{isotropic} uses a direct measurement of
the two-point correlation function of the $\hat p_{ik}$ values using
\textsc{TreeCorr}\footnote{\url{https://github.com/rmjarvis/TreeCorr}} \citep{Jarvis04, TreeCorr}.
This is the default \code{optimizer},
\edit{since it is typically much faster and may be somewhat
more accurate than the the maximum likelihood method.}
\item
\code{anisotropic} is similar to \code{isotropic} except that it uses the
\code{TwoD} binning option in \textsc{TreeCorr}, which measures the correlation
function in two spatial directions, rather than just radially.
This is particularly useful if the PSF pattern is
significantly anisotropic (e.g. due to a predominant wind direction).  In this
case, the resulting kernel is anisotropic as well.
\item
\code{none} does no optimization of the hyper-parameters.  This is probably
only useful for tests where you may know the true kernel and don't want any
optimization to be performed.
\end{enumerate}

Typically when Gaussian processes are used on noisy data, one should include
a ``white noise kernel'' representing
noise in the individual measurements $\hat p_{ik}$.
\edit{\piff\ (by default) accounts for this using the estimated variance from the model fits.}
All models in \piff\ calculate variance
estimates along with the best fit values, which are usually sufficient for this purpose.
However, there is an option to include an additional white noise kernel if one thinks that these
estimates are insufficient to completely describe the measurement noise.

Gaussian process regression makes the assumption that the expectation
value of the parameters $p_{k}$ at each location is zero
\edit{(or some function that can be modeled with hyperparameters)}.
If this is not true,
e.g. because there is some static pattern imposed on top of the Gaussian
variation, then one would need to subtract off this static pattern before running
the Gaussian process interpolation.  \piff\ includes a mechanism to do this,
called \code{meanify}, which computes the mean values of the measured
parameters over a number of exposures to estimate the static pattern.
This can then be subtracted before using the Gaussian process interpolation.

In simulations of atmospheric PSF variation, we have found that Gaussian
process interpolation works very well when the true PSF only includes an
atmospheric component.  \fig{fig:gp} shows an example where we simulated
a purely atmospheric PSF pattern.  The first panel shows the true pattern of
PSF ellipticities without noise.  
The second shows the ellipticities with measurement noise.  
We fit this pattern both with a second order polynomial across each CCD
and with a Gaussian process (using the \code{anisotropic} optimizer) across
the full FOV.  
In both cases, we reserved 20\% of the stars to use
for validation.  The residuals on the reserve stars in the two cases are shown
in the third and fourth panels.  The Gaussian process is clearly superior,
leaving much smaller residuals than the polynomial, 
\edit{although to be fair, this is experiment is slightly circular, since the simulation
was made with a Gaussian process for the atmosphere.}

We have not used this method for DES Y3
data though, because it does not work very well when the real PSF includes
an optical component, including discontinuities at the chip boundaries.  
We are still working on a PSF mode that can properly
account for optical effects (including the chip boundaries) coupled
to a Gaussian process interpolation for the atmospheric component
(cf. \sect{sec:future:composite}).

\subsection{Basis-function polynomial interpolation}
\label{sec:basis}

An alternative to the simple polynomial interpolation implemented in \code{Polynomial}
is to delay the full solution of the $p_k$ values at the location of each star until the
code is ready to also fit for the interpolation coefficients.  This scheme is implemented
in the \code{BasisPolynomial} class.  This is the interpolation scheme we used in the
DES Y3 WL analysis to interpolate the \code{PixelGrid} model.

For the previous interpolation schemes, the maximum-likelihood
estimate $\mathbf{\hat p}_i$ is derived using an iterative process,
since the model \eqn{dmodel} is not linear in the fluxes and centroids of
the stars, and may not be linear in the parameters $p_{ki}$.  Furthermore
it is important to note that the likelihood in \eqn{likelihood}
contains the model in the noise estimate.  The iteration step consists
of finding the least-squares solution for the differential parameter
shift $\delta\mathbf{\hat p}_i$ to the equation
\begin{linenomath*}
\begin{align}
\label{eq:design}
  \mathbf{A}_i \delta \mathbf{\hat p}_i & = \mathbf{b}_i  \\
  A_{i\alpha k} & \equiv \frac{\partial d_{i\alpha}}{\partial p_{ik}}
                  \left(\sigma^2_{i\alpha}{+}d_{i\alpha}\right)^{-1/2}
  \\
  b_{i\alpha} & \equiv \left(\hat d_{i\alpha}{-}d_{i\alpha}\right)\left(\sigma^2_{i\alpha}{+}d_{i\alpha}\right)^{-1/2}
\end{align}
\end{linenomath*}
\piff\ computes $\mathbf{A}_i$ and $\mathbf{b}_i$ of the design
equation using the values of $\mathbf{\hat p}_i$ and $d_{i\alpha}$
derived in the previous iteration.  The derived
$\delta\mathbf{\hat p}_i$ are then added to the solution
to obtain the parameters for the next iteration.
After the first
pass, the shifts are generally relatively small adjustments in the fit.

For the \code{BasisPolynomial} interpolation, we instead
directly model $\mathbf{p}_i$ in terms of the
interpolation coefficients $Q_{km}$ via
\begin{linenomath*}
\begin{equation}
p_{ik} = \sum_{m} Q_{km} K_{im}(u_i, v_i, \dots)
\end{equation}
\end{linenomath*}
where $\mathbf{K}_{i}$ is a basis vector giving the relevant polynomial terms
at the location of star $i$: ${\mathbf K}_i=\{1, u_i, v_i, u_i^2, u_i v_i, v_i^2, \dots\}$.
The design equation for the iteration step of an individual star can
be rewritten as
\begin{linenomath*}
\begin{equation}
\mathbf{A}_i \delta \mathbf{Q} \mathbf{K}_i = \mathbf{b}_i
\end{equation}
\end{linenomath*}
The \code{BasisPolynomial} algorithm essentially concatenates the
design equations for all the stars $i$ into a single system which can
be solved for $\delta\mathbf{Q}$, thus using the information from all
of the stars at once.

One advantage of the \code{BasisPolynomial} interpolation scheme over the simpler
\code{Polynomial} scheme is that it is less affected by missing data such as from
cosmic rays or bad columns.  The design matrix for a star with missing data can
omit those pixels from the constraint, in which case the solution for the coefficients $Q_{km}$ 
will use the other stars to constrain any degrees of freedom that involve
those pixels.  This feature is especially important for the \code{PixelGrid} model, where
missing data generally leads to singular (or at least poorly conditioned) matrices for those stars.
With an interpolation scheme that requires the solution $\mathbf{\hat p}_i$ be completed
for each star, these stars would have to be excluded.
But using \code{BasisPolynomial}, such stars can still provide the information
that is available, while leaving out the pixels that cannot provide any constraining power.

Another nice feature of this scheme is that it is straightforward to properly include
the uncertainties.  This scheme directly uses the variance
in the pixel counts without requiring the intermediate calculation of 
$\mathrm{Cov}\left(\mathbf{\hat p}\right)$ and propagation of that through the
interpolation fit.

\section{Solving for the PSF}
\label{sec:psf}

When solving for the full PSF solution, \piff\ uses an iterative approach to
successively improve the solution.  This has a number of advantages over
a single-pass direct solution.  First, it allows for the possibility of rejecting
outlier stars after each iteration if there are some stars that are not good
exemplars of the PSF (perhaps because they are binary or have a 
neighbor contaminating the image around the star).  Second,
it makes it easier to handle missing data (e.g. bad pixels or
columns); we may, if we wish, use the solution from the other stars
to backfill the missing pixels for a particular star,
\edit{although this is not necessary when using \code{BasisPolynomial}}.
Third, as noted
above, the iterative process allows us to use the model $d_{i\alpha}$
rather than the measurement $\hat d_{i\alpha}$ to estimate each
star's Poisson contribution to the noise. Using the Poisson
shot noise estimated from the observed signal is 
necessarily biased, since pixels that happen to scatter high will have
too high an estimated variance and be under-weighted, and those that scatter low will have
too low an estimate and be over-weighted.

The overall procedure starts by making a small cutout image of the
observed pixels from the full image around each star,
centered as nearly as possible at the star's input position.
By default these cutout images are 32 $\times$ 32
pixels square, but this can be changed if desired.  There are options
to remove some stars at this point based on measurements of the
star on these cutout images.

Then for each input star, \piff\ initializes the solution with a rough guess
for the model.  Different models do this in slightly different
ways, but generally they start with something like a Gaussian 
profile matched to the stars' measured second moments.  This defines the initial
model vectors $\mathbf{\hat p}_i$.

During each iteration of the solution phase, \piff\ either solves for 
the shifts $\delta\mathbf{\hat p}_i$ according to the chosen model,
as described in \sect{sec:model}, or
computes the design equation for these shifts (\eqn{eq:design})
if using \code{BasisPolynomial} for the interpolation.

Then \piff\ solves for the interpolation coefficients according to the
chosen interpolation scheme, as described in \sect{sec:interp}.  
This finishes the solution for this
iteration.  The solution is evaluated at the location of each input star
as the starting point for the next iteration.

At the end of each iteration, \piff\ can look for outliers and remove them
from consideration for the next pass.  It also computes the overall
$\chi^2$ and degrees of freedom (dof) of the solution.  
If there were no outliers removed and the
change in $\chi^2$ is below a user-set threshold or if it has reached
a user-set maximum number of iterations, then the process stops,
and the solution is written to an output file.

\subsection{Input data}
\label{sec:input}

\piff\ has a number of options for specifying the input data.
First, it needs the image (or images)
making up the field for which to solve for the PSF.  
One can run \piff\ on a single CCD image or on all the images from a
single exposure (or some fraction thereof).
Each image is typically
input as a FITS file, including a weight or variance map, the world coordinate
system (WCS), and possibly a bad pixel mask.  The user can specify
specific bits in the bad pixel mask to exclude from consideration.

One can also specify a different WCS to use rather than the one in
the FITS file if desired.  We used this feature in the DES Y3 WL analysis
to use the improved \pixmappy\ solution (cf. \sect{sec:wcs}), which includes tree rings
and other subtle effects that cannot be expressed in the regular
FITS WCS specification.

The other required input is a list of stars and their positions on the
image(s).  These can be given either as $x$ and $y$ pixel values or as
right ascension and declination.  \piff\ does not currently have any
ability to determine which objects are stars automatically, but it can
use a flag column to select only a portion of the given input catalog
if appropriate.

There are also options to remove some stars based on features
of their images.  For instance, one may specify a saturation threshold
to exclude stars that have any pixels above this value.
If the star's cutout image is partially off the image, it is normally excluded, but
one can optionally keep such stars.
One can also
exclude stars whose measured size is an extreme outlier compared
to the other stars (e.g. due to neighbors or image artifacts) or
whose measured signal-to-noise ratio (SNR) is smaller than a given value.
For the DES Y3 WL analysis, we excluded stars with SNR $< 20$.

In addition to removing stars that are deemed bad for some reason,
there is also an option to reserve some stars from being used for the
PSF solution to serve as fair test stars for diagnostics.
These stars are not used for any part of the iterative solution, but
they are included in the output catalog,
marked with a flag indicating that they are reserve stars.
For DES Y3, we reserved 20\% of the input stars.

The weight map image is used to determine the measurement noise
on the pixel values.  The weight map is typically the inverse variance
of each pixel, but \piff\ can also read a (not-inverse) variance map.
The input weight should ideally include only
the estimated variance from the sky, read noise, dark current, etc.,
not including the Poisson shot noise of the signal itself.
(This is the case for the processed images produced by DES
data management.)
As noted above, estimating the source shot noise from the measured
counts will induce a bias on the measured PSF and fluxes.
If the input weight (or variance) map includes the variance of the
signal as well as the uniform sources of noise, then \piff\ has an option
to remove it using the gain (either provided or fit for by \piff\
\edit{from the variance map and the image}).

Finally, another option that we found to be important in the DES Y3 WL
analysis is to down-weight the brightest stars.  If each
star keeps its real noise estimate, then most of the weight comes from
just the few brightest stars in the image.  This tends to bias the
interpolation solution to overfit the modes that pass through these
stars.  Therefore, we have an option to effectively limit the nominal SNR
of the bright stars to a given maximum.  \piff\ doesn't actually
add noise to achieve this nominal SNR; rather, it just decreases
the weight map to the level that would give bright stars this SNR value.
The default for this parameter is 100, which was used for
DES Y3.  This choice was found to produce good results
(cf. \sect{sec:tests}),
but the results were not very sensitive to changing this by a factor of 2 or so.

Note that while Piff can be used on co-added images, rather than
single-epoch images, it is not recommended if the
dithering strategy includes large offsets.  The PSF in the co-add is
discontinuous at the location of every chip edge from the single-epoch
exposures.  These cause problems when trying to interpolate the
PSF across the co-add image.
\edit{Additionally, the current implementation
assumes the noise is uncorrelated across pixels, which would not be true
in general for co-added images.}

\subsection{Outliers}
\label{sec:outliers}

At the end of each iteration, there is an option to remove stars that are determined
to be outliers and thus are probably not good exemplars of the PSF.  
Currently there is only one outlier method, although the code is written to
accommodate the addition of other algorithms for identifying stars to remove.

The \code{Chisq} outlier method looks for stars whose measured $\chi^2$
is very large.  Specifically, for each star $i$, it sums over pixels
$\alpha$ in the cutout of the star:
\begin{linenomath*}
\begin{equation}
\chi_i^2 =  \sum_{\alpha \in i}
                                           \frac{\left(\hat d_{i_\alpha} {-}
                                           d_{i\alpha}\right)^2}{
                                         \sigma_{i\alpha}^2 {+} d_{i\alpha}}
\end{equation}
\end{linenomath*}
where the sum is over all of the pixels in the cutout image for that star.
As discussed above, the total noise in each pixel is taken to be the sum
of the read/background variance $\sigma_{i\alpha}^2$ and the Poisson
variation of the expected counts, $d_{i\alpha}$.
If the $\chi^2$ value is more than some threshold, then the star is removed
from consideration for subsequent iterations.

The usual way to specify the threshold is in terms of an effective number of ``sigma''.
Given a specified \code{nsigma} value, \piff\ calculates the corresponding probability that a
Gaussian distribution could exceed this many sigma.\footnote{It uses the two-sided 
probability, $p\,{=}\,\mathrm{erfc}(\code{nsigma}/\sqrt{2})$.  This is merely a shorthand to
allow users to convert intuitive ideas of ``n sigma'' into a probability.  It doesn't actually
refer to some number of any meaningful ``sigma''.}
For instance, \code{nsigma=2} corresponds to $p\,{=}\,0.05$, \code{nsigma=3} corresponds to
$p\,{=}\,0.003$, etc.
If preferred, users can also input the probability directly.

Then for each star, \piff\ calculates the threshold for which this is the probability that the measured
$\chi^2$ would exceed the threshold purely from statistical noise,
given the number of degrees of freedom for that star\footnote{
We use the usual definition that the number of degrees of freedom is
the number of data points minus the number of model parameters;
however, we note that this definition is not necessarily correct or meaningful for nonlinear
models \citep{Andrae10}.  We believe it is acceptable for this purpose, given the kinds of models that
are used for PSF modeling though.
}.
For the DES Y3 WL analysis, we used \code{nsigma=5.5}, which corresponds to $p= 4 \times 10^{-8}$.

One can also specify a maximum number of stars to reject in each iteration.  This is generally
a good idea, since a small number of outliers can potentially skew the solution to the point where
almost all of the stars have a bad $\chi^2$ value.  For DES Y3, we limited the
rejection to at most 1\% of the stars in the exposure (rounded up), 
which typically translated into either 1 or 2 stars per iteration.

\subsection{Output files}
\label{sec:output}

Once the iteration has converged, \piff\ writes the final solution into an output FITS
file.  The file format is rather complicated, using many HDUs (header/data units) to store the various kinds
of information in a modular way.  For instance each \code{Model} class and each \code{Interp}
class stores different kinds of information, so each uses one or more HDUs in a class-specific way.

Users do not need to know anything about this file format however, since the \code{piff}
Python module has code to read the output file and reconstruct a \code{PSF} object that 
can compute the correct PSF profile at an arbitrary location.  \piff\ is designed to serve the
roles both of solving for the PSF solution and of using that solution for further analysis.

In addition to the FITS file containing the final PSF solution, one can also choose to have
\piff\ generate a number of diagnostic output files.  These include a number of plots
including residuals as a function of position in the field, diagnostic statistics, and more.
These plots are not very useful for characterizing the quality of the fits for a large data
set (such as DES Y3), since they are made for one exposure at a time, but they can
be very useful for simple sanity checks when trying out different configuration choices
for a particular data set.

One can also have \piff\ output a catalog with
measurements of the size and shape of both the PSF model and the actual star
images.  These catalogs are more useful for large scale diagnostics, since one can
generate statistics for many images by combining these data.  The plots in
\sect{sec:tests} are made from these residual measurements.\footnote{Technically,
the measurements for those plots were done outside of \piff, but the code we had
been using to generate those measurements has since been ported into
\piff\ to make it easier to generate such data in the future.}

\section{Data} 
\label{sec:data}

The first three years of data from the Dark Energy Survey (DES Y3) covers nearly 5000 square degrees
of the (mostly) southern sky, and includes close to 40,000 exposures reaching an $i$-band limiting magnitude
\edit{(10$\sigma$ detection, $2"$ aperture)} of 23.4 \citep{des-dr1}.
We refer to \citet{y3-gold} and \citet{imageproc} for most of the details about the data reduction, including
flat fielding, sky-subtraction, noise characterization, and object detection,
but we mention some relevant points here.

\subsection{Brighter-fatter correction}
\label{sec:bf}

One of the most important improvements in the data reduction process compared to the DES Y1 reduction
is that a correction was applied to remove the ``brighter-fatter effect'' \citep[BFE;][]{Antilogus14, Guyonnet15}.
The BFE is a natural consequence of Coulomb's Law for the electrons being accumulated
in the detector.  
The electrons accumulated in high-flux pixels repel some of the other electrons
arriving later in the exposure, which would have been expected to fall in these pixels.
The later-arriving electrons tend to be pushed outward from the centers of bright objects, causing
these objects to appear somewhat larger than they would have appeared in the absence of BFE.
The enlargement is more significant for brighter and more compact sources (i.e. the ones with the
highest surface brightness).

The impact of BFE is quite significant in DECam images, noticeably affecting the sizes of
the brightest three magnitudes of stars on any given image \citep{Melchior15, sv-shearcat, y1-shearcat}.  In the Y1
analysis, we were thus forced to remove these stars from our sample of PSF stars to avoid
biasing the PSF size.  

The correction procedure applied to the Y3 images was
originally proposed by \citet{Antilogus14} and quantified for DECam by
\citet{Gruen15}.  It involves moving some of the flux observed on the image back to where it would have
fallen in the absence of BFE.
This correction is applied directly to the pixel values early in the data reduction process \citep{imageproc}.  

As we will see below (\sect{sec:mag}), the correction does not work perfectly, but it corrects for
about 90\% of the full effect.
We found that there was still a non-negligible bias in the sizes and shapes of
stars within 1.2 magnitudes of saturation.  Therefore, we still needed to remove these very bright stars from our
PSF sample before running \piff. However, this means that we included almost two magnitudes more
stars in our PSF sample than in Y1, which significantly helped improve the Y3 PSF solutions.

\subsection{World Coordinate System}
\label{sec:wcs}

Another important detector effect seen in earlier DES analyses is a circular pattern most easily
seen in the flat field images.  
This effect, known as ``tree rings'', is due to lateral electric fields in the CCDs due to
impurities in the silicon.
The impurities are deposited as the silicon crystals are grown, which leads to a ring pattern
around the initial crystal seed location.
\citet{plazas} showed that the tree rings are primarily an astrometric effect causing the effective
pixel size to vary radially around the centers of the silicon crystals.  

For the DES Y3 WL analysis, we used an astrometric solution that included the tree ring
information as part of the model, fit from the positions of objects on overlapping images \citep{pixmappy}.
The model also included other astrometric effects including the telescope distortion pattern,
edge distortion at the edges of each CCD, and adjustments in the precise positions of each
CCD, which were determined to move slightly each time the camera warmed up.  

The full solution to the world coordinate system (WCS) is implemented by the
\pixmappy\footnote{\url{https://github.com/gbernstein/pixmappy}} software package.
We used this WCS solution for each exposure rather than the simpler TAN-PV WCS solutions stored
in the FITS files.  Since \piff\ models the PSF in sky coordinates,
the astrometric effects of the tree rings and other distortion effects
are accounted for by transforming the image data of each star to sky coordinates.
The led to a significant improvement in the PSF shapes compared to using the native
FITS-based WCS solutions.

\subsection{Selection of PSF Stars}
\label{sec:stars}

\begin{figure}
\includegraphics[width=\columnwidth]{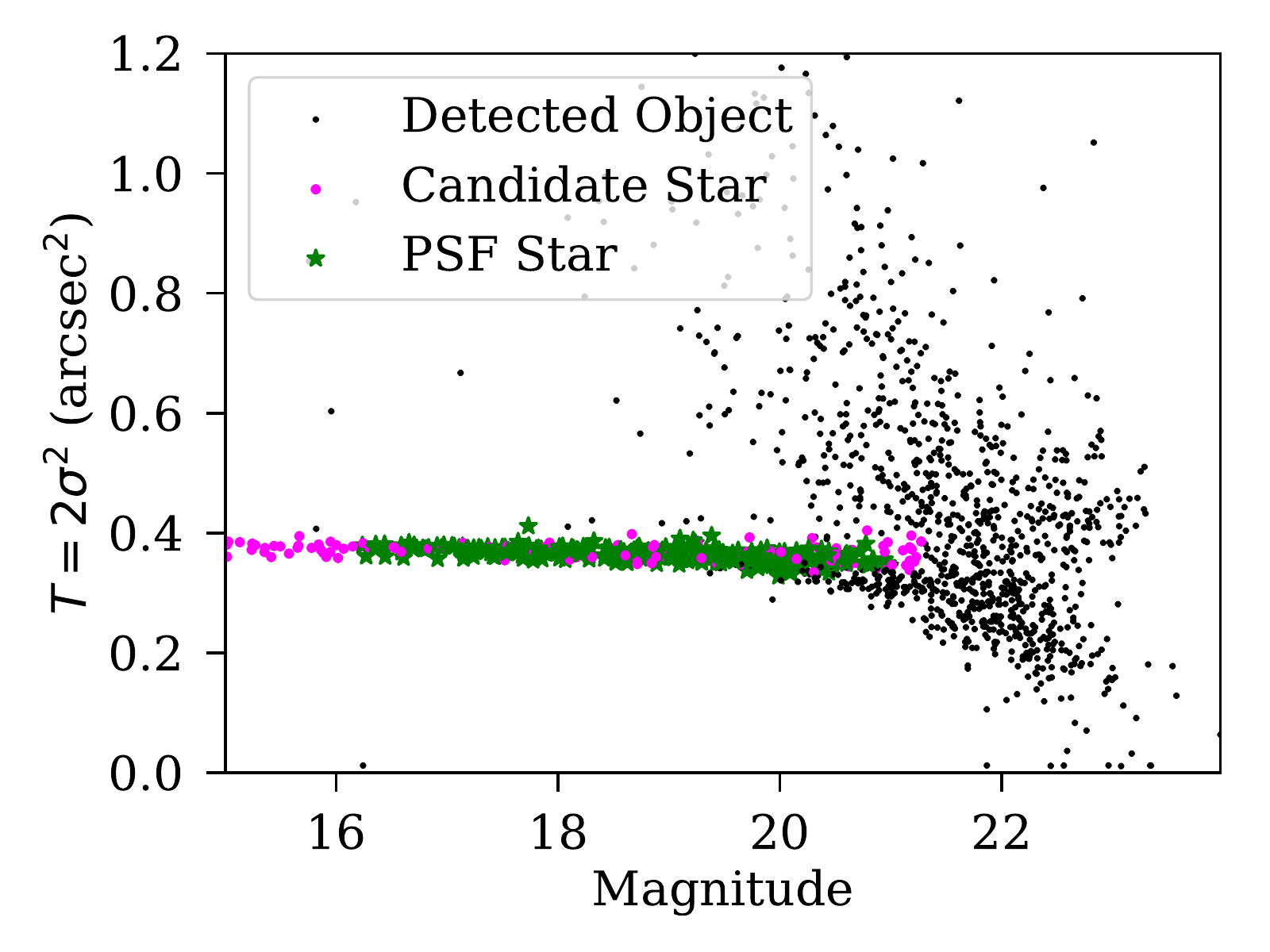}
\caption{
An example size-magnitude diagram for a single CCD image,
used to identify stars.  The size $T=2\sigma^2$ is based on the scale size
of the best-fitting elliptical Gaussian.
The pink and green points are the objects initially identified as stars.
The green points are the ones that pass our selection criteria outlined in \sect{sec:stars},
most notably the
magnitude cut to avoid objects contaminated by the brighter-fatter effect.
These objects are then used to constrain the PSF model.
\label{fig:findstars}
}
\end{figure}

The selection of stars to use for PSF modeling in DES Y3 used the same algorithm as has been used in 
both of the previous
DES analyses: Science Verification \citep{sv-shearcat} and Y1 \citep*{y1-shearcat}.  We refer to those papers
for details, but we provide a brief summary here.

For the initial catalog of objects, we used \sex\ \citep{BertinSExtractor1996}.
We then used a size-magnitude diagram to identify stars as a locus of points with constant size separating from the
larger cloud of galaxies.  
For the magnitude, we used the \sex\ measurement \code{MAG\_AUTO}. For the size, we used the scale size, $\sigma$,
of the best-fitting elliptical Gaussian profile using an adaptive moments algorithm.
The locus is easy to identify by eye at bright magnitudes.  
\fig{fig:findstars} shows an example size-magnitude diagram for a representative DES Y3 image.
The pink and green points were identified as stars, and the black points are other objects
(both galaxies and stars that may be too faint or noisy to be identified as such).

The algorithm we used to automate this identification starts with the brightest 10\% of the objects (excluding
saturated objects) and finds a tight locus at small size for the stars 
and a broad locus of galaxies with larger sizes.
Then the algorithm proceeds to fainter magnitudes, building up both loci, until the stellar locus and the 
galaxy locus start to merge. The precise magnitude at which this happens is a function of the seeing 
as well as the density of stars and galaxies in the particular part of the sky being observed. 
The faint-end magnitude of the resulting stellar sample thus varies among the different exposures.
The green and pink points in \fig{fig:findstars} were identified as stars by this algorithm.

As discussed above (\sect{sec:bf}), the initial data processing included a correction for the brighter-fatter effect.
However, we found that the brightest stars still showed some significant size residuals.  We therefore
removed the identified stars within 1.2 magnitudes of saturation to avoid these stars 
biasing the inferred PSF.  We also removed stars that were close to the edge of the CCD or near 
the DECam ``tape bumps'' \citep{DECam}.  
We also removed a random 20\% of the remaining stars as ``reserve'' stars for the
diagnostic shown below in \sect{sec:tests}.
The pink points in \fig{fig:findstars} show the
stars that were removed for one of these reasons.  
The green points represent the final
stellar sample for this CCD, which was input into \piff.

\subsection{Blacklist}
\label{sec:blacklist}

Immediately upon completion of a \piff\ model, we perform some basic sanity checks to make sure the model
seems plausible.  If a PSF model for a particular CCD is considered suspect for any of the following reasons, 
we enter it into a ``blacklist'' and exclude this CCD from any further analysis.

\begin{itemize}
\item
Too few stars: We flag images for which fewer than 25 stars survived the outlier rejection.
\item
Too many stars: We flag images where more than 30\% of the objects were considered stars.
This is unusual and generally indicates a problem with the star selection, rather than a truly dense stellar field.
\item
Outlier size: If the mean size of the PSF solution for one CCD was very different (4 sigma outlier) from the others
on the same exposure, we assume either the fit or the star selection for that image failed.
\item
Large spread in the PSF model sizes: If the standard deviation of the sizes of the final PSF stars is more
than 20\% of the mean size, then this tends to indicate a bad PSF solution.
\item
Errors when running \piff\ or the stellar locus codes: These were rare but happened occasionally.
\item
No \pixmappy\ WCS solution: A few exposures were taken during periods
with insufficient calibration information to produce a
reliable \pixmappy\ WCS solution.
\end{itemize}

A little fewer than half the exposures had at least one CCD blacklisted by these checks.  Of these,
the average number of CCDs removed from consideration of the downstream analysis was close
to two.  This blacklisting procedure thus led to the removal of about 2\% of the Y3 data.

\subsection{Tests of Stellar Purity}
\label{sec:purity}

\begin{figure}
\includegraphics[width=\columnwidth]{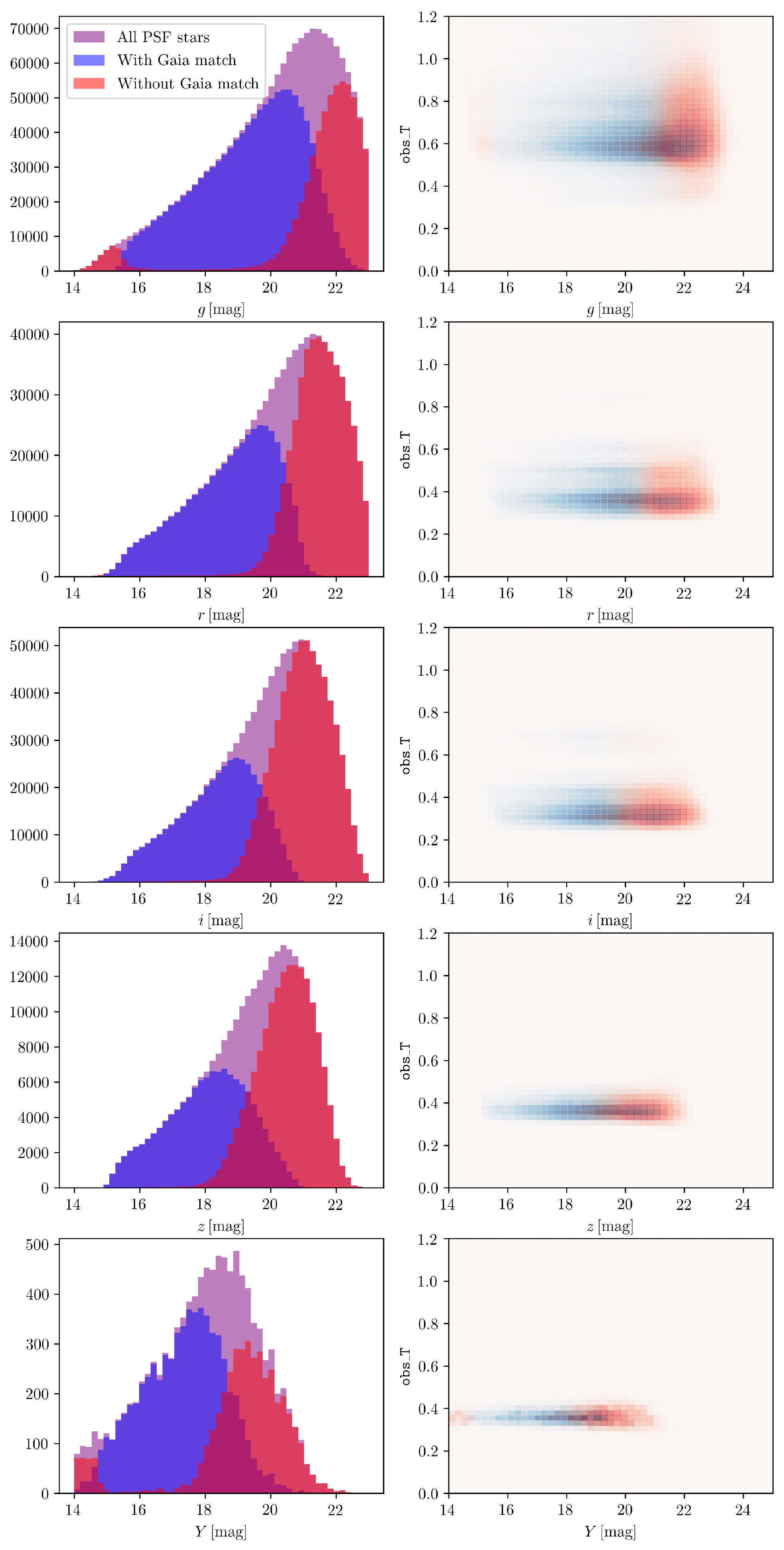}
\caption{
On the left are histograms of the PSF stars  by magnitude for each band: $g$, $r$, $i$, $z$, and $Y$.
The blue region are the PSF stars that were also in the Gaia DR2 catalog.  Red indicates objects that we
consider stars, which are not in the Gaia catalog.  (Purple is all PSF stars.)  On the right are plots showing the density of objects
in size-magnitude space, using the same color scheme.
\label{fig:gaia}
}
\end{figure}

\begin{figure}
\includegraphics[width=\columnwidth]{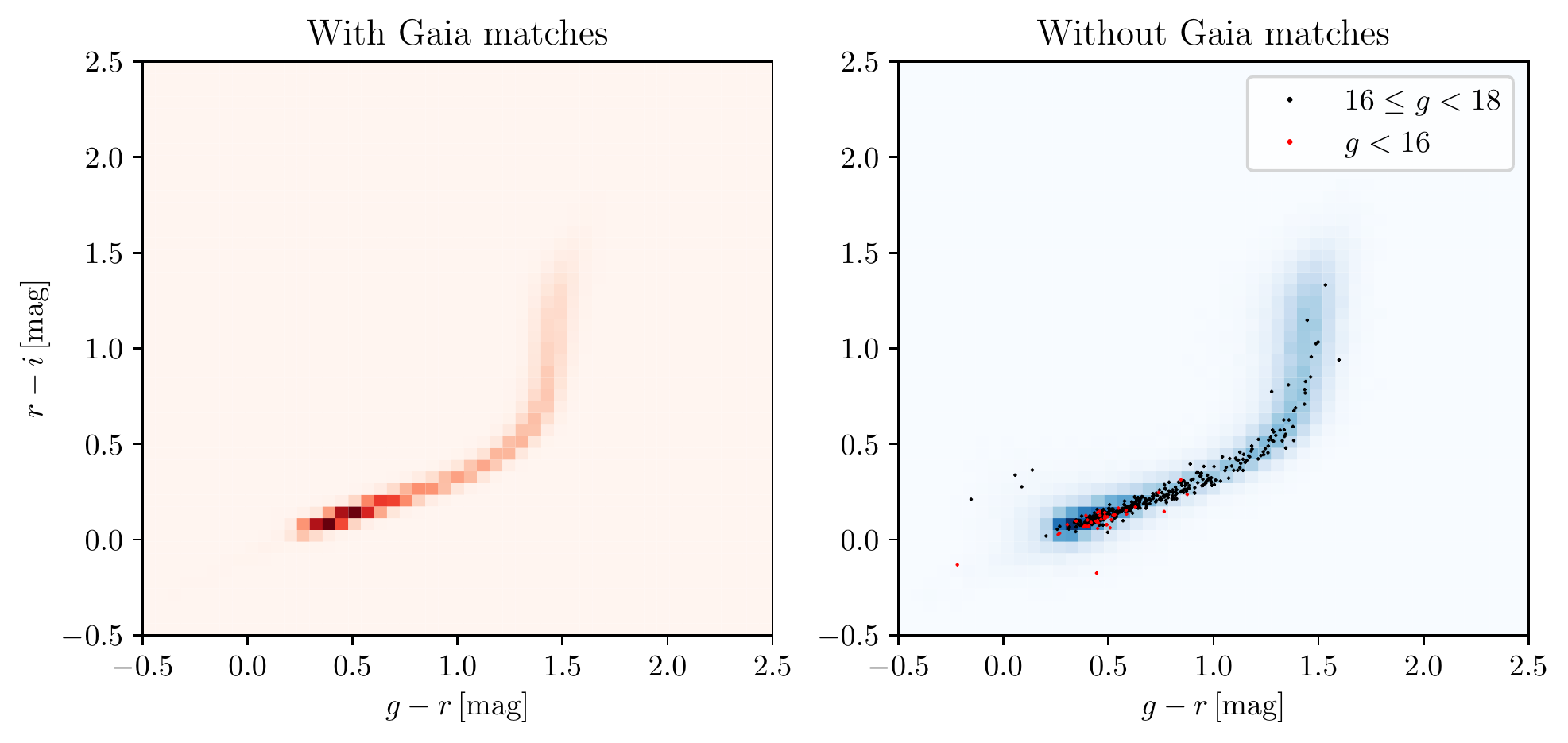}
\caption{Color-color plots of the samples of DES PSF objects which do (\emph{left}) and 
do not (\emph{right}) have matching objects in the Gaia DR2 high purity stellar catalog. 
All objects appear to follow the stellar locus, including the bright unmatched objects shown as individual points
(black for moderately bright and red for very bright).}
\label{fig:gaia_colors}
\end{figure}

To produce an accurate PSF model, the selection of PSF stars should be very pure.
If there are any galaxies included in the sample, then this will bias the resulting PSF size
to be slightly larger than the true size.  We therefore performed two tests to check
whether and to what extent any galaxies were erroneously included in the stellar sample.

First, we follow a test presented in \citet[Appendix C]{amon18} for the Kilo-Degree Survey (KiDS)
to match our catalog of PSF stars to the Gaia Data Release 2 (DR2) stellar catalog \citep{gaia}.
These are believed to be a very pure sample of stars \citep{BailerJones19}, but they do not extend as faint
as our data.  Thus, the matching is only expected to be close to complete at brighter
magnitudes.

The left side of
\fig{fig:gaia} shows histograms of our stellar sample as a function of magnitude
including the portion that matches the Gaia catalog (blue) and the portion that does not (red).
The match is very close to complete at bright magnitudes; more than 98\% of the PSF stars 
match a Gaia star over at least 2 magnitudes in all bands. At the fainter end, the Gaia
catalog becomes incomplete, and the DES data include more stars, which are not matched
to any Gaia stars.  The $g$ and $Y$ bands also show an unmatched population at the bright
end, since some stars brighter than the Gaia sample are unsaturated in DES data.
Within the range of magnitudes where the Gaia catalog is complete, we find that essentially
all the selected PSF stars are matched to Gaia stars. This implies that there are very few galaxies being
included in the sample at these magnitudes. 

In \fig{fig:gaia_colors}, we show the colors of our selected stars with (left) and without (right) matches
in the Gaia DR2 high purity stellar catalog.
The objects without Gaia matches follows the same stellar locus as is seen for the objects with
Gaia matches.  In particular, this is true both for the bright unmatched objects (shown as individual
points) and for the
fainter objects (shown as the blue intensity scaling).  
There does not seem to be any significant sub-population of the bright or faint selected stars
with different color properties than those of the high-confidence Gaia stars.

At fainter magnitudes, where the Gaia catalog is incomplete, we also looked at the size
distribution of the stellar sample.  The right side of \fig{fig:gaia} shows
size-magnitude diagrams of the stars, color coded in the same way as the histograms on the left.
This test is less clear than where a direct match is possible, but there does not
appear to be any significant population of objects with sizes at faint magnitudes that
differ noticeably from the sizes of the bright stars, with the possible exception of $g$ band.
For $g$ band, there does seem to be a cloud of red points with larger sizes than would be
expected from the blue points.  It is thus possible that there are some galaxies being included
in the stellar population for $g$ band.  We chose not to use the $g$ band for any Y3
weak lensing analysis, in part because of this apparent contamination\footnote{
The $g$-band PSF solution was found to have unacceptably large rho statistics
(cf. \sect{sec:rho}), possibly due to this contamination, but also possibly due to
other systematic effects that are strong in the $g$ band such as differential chromatic
refraction (cf. \sect{sec:color}).}.

\begin{figure}
\includegraphics[width=\columnwidth]{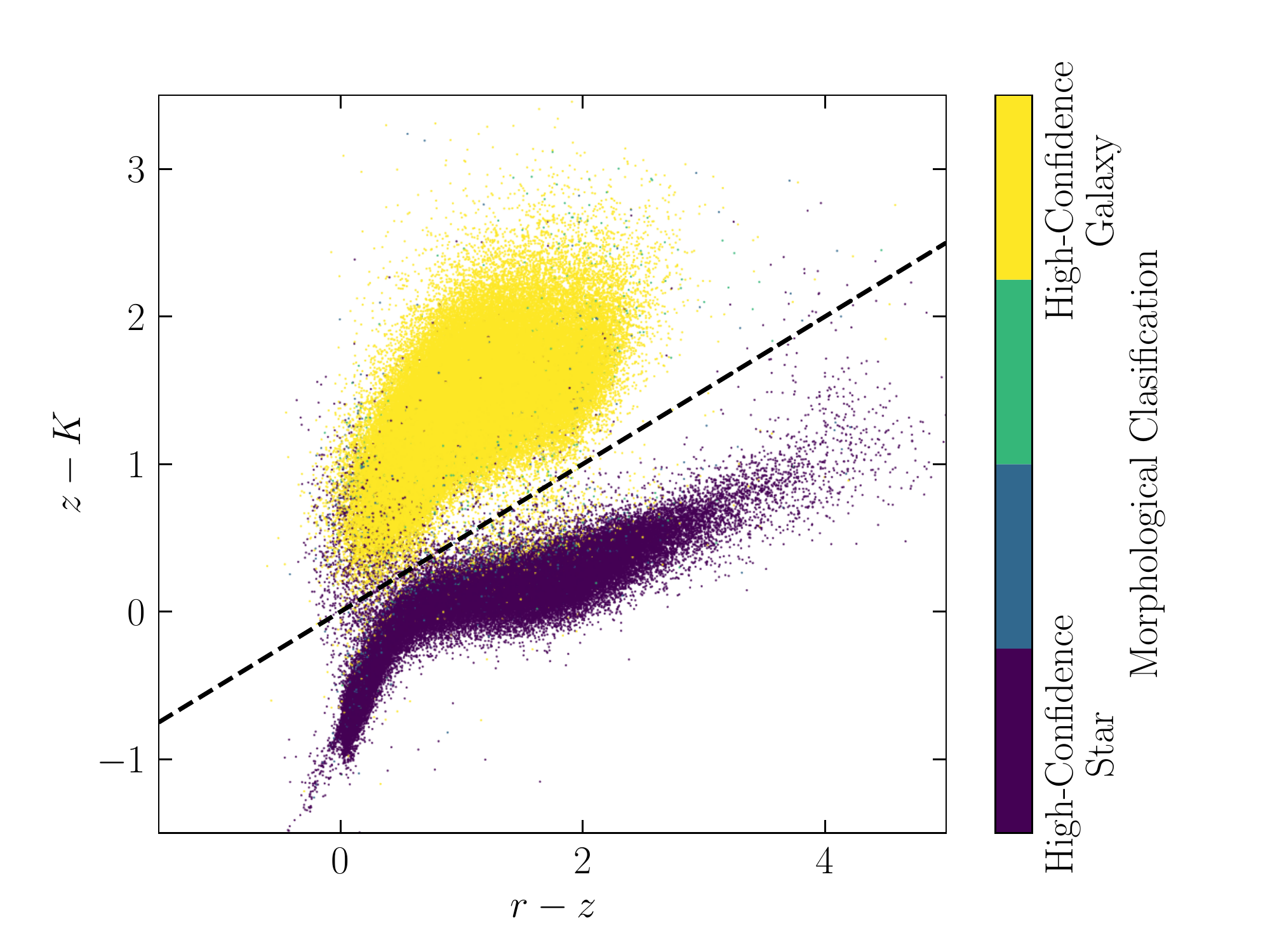}
\caption{
A color-color plot using VHS $K$ band for one of the colors, motivating our VHS-based
color selection test, \eqn{eq:vhs_color_cut}. A random sample of 0.5 million matched DES-VHS objects is plotted.
Morphological classifications indicated by the colorbar come from the Y3 Gold catalog
\citep{y3-gold}.
\label{fig:vhs}
}
\end{figure}

\begin{figure}
\includegraphics[width=\columnwidth]{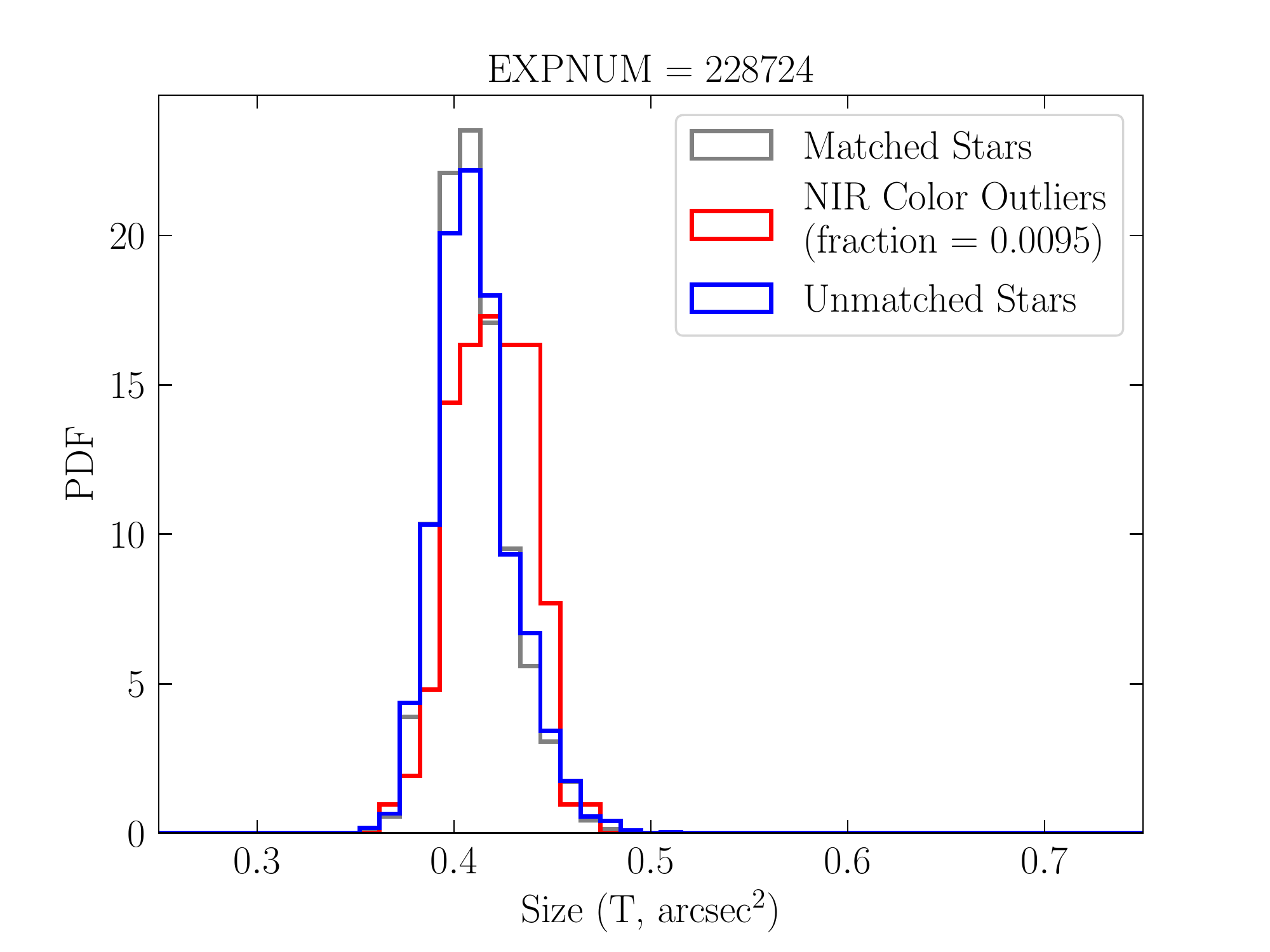}
\caption{
Histograms of the sizes (cf. \eqn{eq:Tdef}) for PSF stars in the VHS overlap region according to whether they were successfully
matched to the VHS catalog, and if so, whether they had colors consistent with being a star (gray) or
not (red).  The unmatched stars are blue.  The outliers show a slightly higher mean size than the
high-confidence stars, but we calculate that this bias is not significant.
\label{fig:vhs_size}
}
\end{figure}

Another test of the stellar purity comes from the infrared (IR) colors of the objects.  In particular, if we include
$K$-band colors from the VISTA Hemisphere Survey \citep[VHS]{vhs}, then we find a simple color selection,
provides a very good discrimination between stars and galaxies:
\begin{linenomath*}
\begin{equation}
(z - K) > 0.5 \times (r - z)
\label{eq:vhs_color_cut}
\end{equation}
\end{linenomath*}
This selection is shown in \fig{fig:vhs}.  We consider the points that appear above this
line (i.e. the $>$ condition in \eqn{eq:vhs_color_cut}) as likely to be non-stellar.  These
``color outliers'' are thus likely to be poor objects to include in the PSF catalog.

Some motivation for the use of IR bands for star-galaxy discrimination was given by \citet{Jarrett2000},
who point out that galaxy light is dominated by old stellar populations with significant flux at 2 microns and 
that their redshifts tends to push additional light into IR bands.  Similar color cuts using a combination of optical 
and near-IR colors have been found to be effective by, e.g., \citet{Ivezic02, Baldry10, y1-sg}.

Unfortunately, the VHS catalog only covers about half of the DES Y3 footprint, 
and 2MASS \citep[The Two Micron All Sky Survey,][]{2mass} and 
WISE \citep[The Wide-field Infrared Survey Explorer,][]{wise}
are both too shallow for this purpose.
Therefore, we cannot apply
this cut to our entire input sample.  Additionally, this idea for selection was proposed after
our Y3 PSF catalogs were finalized, so we did not even apply it for the portion where we had
VHS overlap.  Rather, we use this test to quantify how much the interloping galaxies might be 
biasing the size of the PSF.

\fig{fig:vhs_size} shows the distribution of observed sizes of PSF stars in a single representative
exposure according to their
infrared colors.  Blue shows the stars that were not matched to VHS observations, 
and so do not have a $K$-band magnitude.
Gray shows the matched stars that fall below the condition in \eqn{eq:vhs_color_cut},
and thus are expected to be true stars.  
Red shows the matched stars that fall above this condition, called color outliers,
which are likely
to be galaxies.  For this exposure, 
the color outliers constitute less than 1\% of our PSF stars, and their
mean size is larger than the high-confidence stars by about 4\%.
This means the interloping galaxies may be inducing a fractional bias in the size of about $4 \times 10^{-4}$
for this particular exposure.

\begin{figure}
\includegraphics[width=\columnwidth]{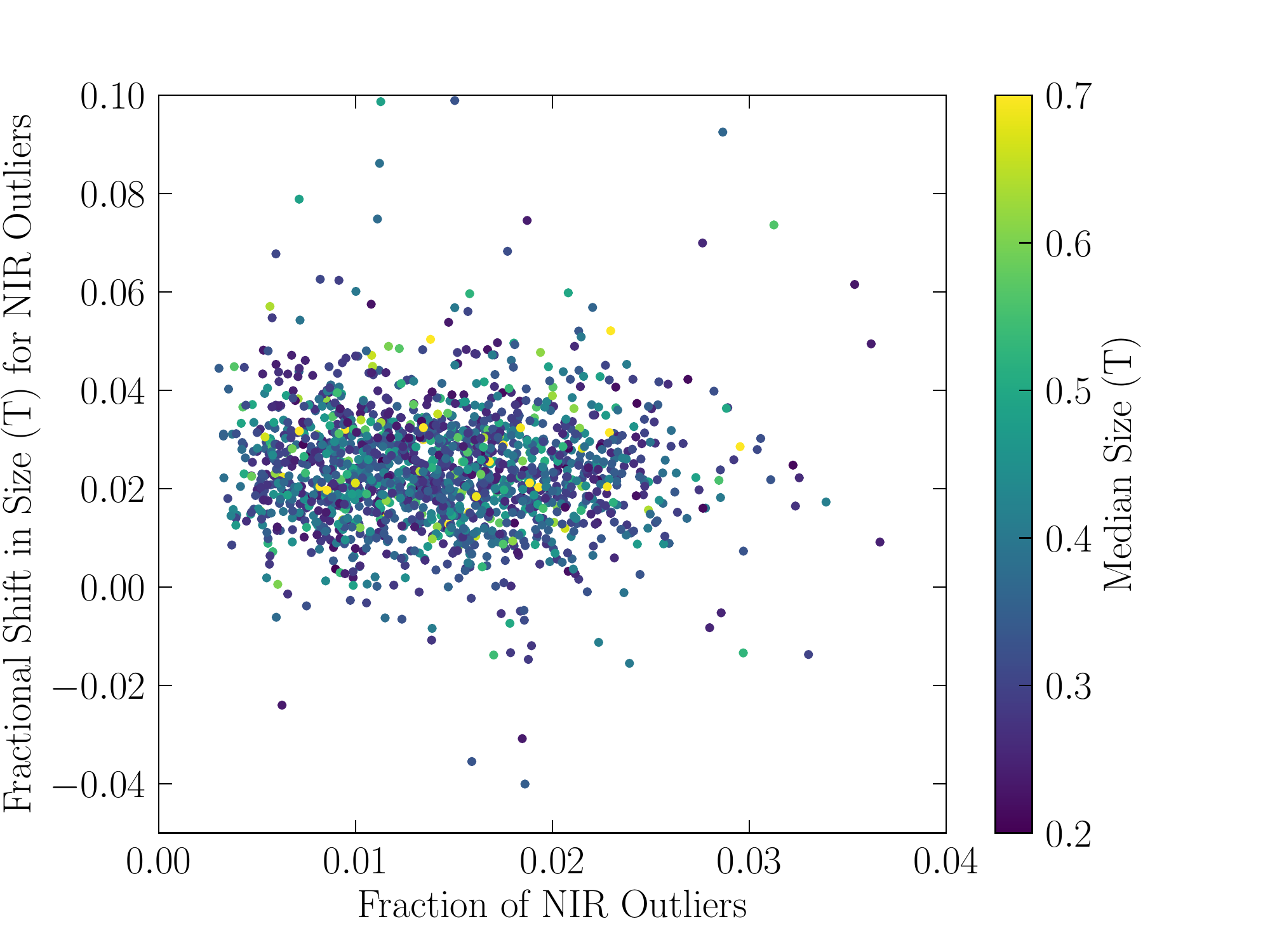}
\caption{
The fraction of identified PSF stars found to be color outliers (cf. \fig{fig:vhs_size}) and their
mean fractional shift in measured size, compared to the stars that fall below the condition in
\eqn{eq:vhs_color_cut} and are thus high-confidence stars. Each point represents a single exposure.
Results are plotted for a random sample of 1685 exposures \edit{in the $i$ band} each having at least 100 matched VHS objects (median of 7000 matched objects).
\label{fig:vhs_all}
}
\end{figure}

\fig{fig:vhs_all} shows the fraction of color outliers and their mean fractional shift in measured size,
$\langle \Delta T\rangle / \langle T \rangle$, 
for a random sample of Y3 exposures \edit{in the $i$ band} with matching VHS data.
While the specific values vary somewhat from exposure to exposure, there is no evidence
of any exposures with particularly bad stellar identification.  Nearly all exposures have less than
3\% outlier fraction with a fractional size shift less than 0.05.  This implies that the
fractional bias in the PSF size is nearly always less than $1.5 \times 10^{-3}$.
This bias is small enough not to be important for Y3 weak lensing analyses, but it is not completely
negligible, so we will try to improve upon this for the Y6 PSF modeling.

\begin{figure*}
\includegraphics[width=\textwidth]{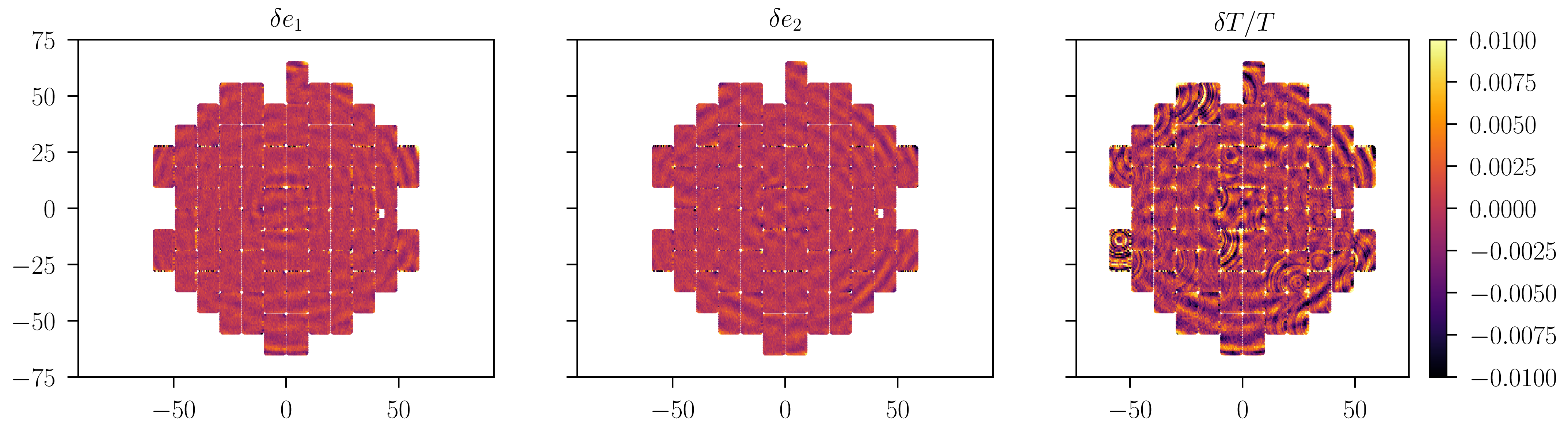}
\caption{
Maps of the two components of the residual shape ($\delta e_1$ and $\delta e_2$) and the 
fractional residual size ($\delta T/T_\mathrm{psf}$)
as a function of position in the focal plane for $riz$ bands.
The shape components have a slight radial pattern, corresponding to high spatial
frequency modes
that our 3rd order interpolation across each CCD was unable to completely model.  The size residuals show noticeable
tree ring patterns in many CCDs.  These patterns are much smaller with the \pixmappy\
astrometric solution than when we use the native FITS WCS solutions, but they are still
significantly non-zero.
\label{fig:fov2}
}
\end{figure*}

\section{PSF Diagnostics}
\label{sec:tests}

We have performed a number of diagnostic tests of the quality of the DES Y3 \piff\ solutions.
Among the various uses of the PSFs, weak lensing shear estimation
generally has the strictest requirements on the quality of the PSF estimates.
We have therefore primarily focused our attention on tests of errors in the PSF solutions that 
could adversely impact weak lensing shear estimates.

All of the diagnostics are calculated using a set of ``reserve'' stars, which were not
used to constrain the PSF solutions.  
We reserved 20\% of the selected stars (cf. \sect{sec:stars}) at random to constitute
this reserve set and removed them from the list of stars passed to \piff.  
These stars thus provide
an unbiased estimate of the errors at random locations in the image.

We calculated the PSF solution at the location of each of these reserve stars to compare
to the actual observed surface brightness profile of the star.
We used \ngmix\ \citep{ngmix} to measure the second moments of both the observed 
stars and the \piff\ models drawn at these locations: 
\begin{linenomath*}
\begin{align}
I_{uu} &= \int I(u,v) u^2 du dv\\
I_{uv} &= \int I(u,v) u v du dv\\
I_{vv} &= \int I(u,v) v^2 du dv
\end{align}
\end{linenomath*}
These moments are not computed by direct summation, which is somewhat unstable in practice.
Rather \ngmix\ finds the best fitting elliptical Gaussian profile to the observed flux distribution.
The above moments are then taken to be the analytic second moments of this profile.

The size $T$ is defined as the trace of the moment matrix,
\begin{linenomath*}
\begin{equation}
T = I_{uu} + I_{vv},
\label{eq:Tdef}
\end{equation}
\end{linenomath*}
and the complex ellipticity can be calculated as
\begin{linenomath*}
\begin{equation}
e = \frac{I_{uu} + 2i I_{uv} - I_{vv}}{I_{uu}+I_{vv} + 2 \sqrt{I_{uu} I_{vv} - I_{uv}^2}}
\label{eq:edef}
\end{equation}
\end{linenomath*}

\subsection{Residuals in the field of view}
\label{sec:fov}

\fig{fig:fov2} shows the residuals of the the shape ($e_1$ and $e_2$) and size ($T$) measurements
for the reserve stars in $riz$ bands as a function of position on the DECam focal plane.  All three show a noticeable
oscillatory pattern consistent with a 4th order polynomial on each CCD.  This is due to the fact that our
interpolation scheme is only at 3rd order, so the smallest order of variation not captured by our
PSF model is at 4th order\footnote{We tested using 4th order polynomials for the solutions, and other
diagnostics, such as the rho statistics (cf. \sect{sec:rho}), became worse, probably due to over-fitting
in fields with smaller numbers of stars.}.

The size residuals show \edit{some additional, much smaller,} circular patterns, which are more prominent on some chips than others.
The most obvious example of this is found in the lower of the two left-most
CCDs, but it appears quite significantly on several others as well.  These patterns are very similar to the tree-ring
patterns in the astrometry, implying that our procedure of measuring and interpolating the PSF in sky coordinates
was not sufficient to fully remove the effects of the tree rings on the PSF size.

We investigated switching to using chip coordinates rather than sky coordinates to model the PSF, and the 
tree ring patterns in the residuals became significantly worse.  The same was true when we used the simpler WCS
solutions in the FITS files rather than using \pixmappy.  Thus we know that using the \pixmappy\ WCS is working
to reduce the impact of the tree rings; it just isn't sufficient to
fully remove all of the effect.  

We believe that at least part of the reason for this is that 
the total PSF size includes a component due to electron diffusion in the CCD.
This component of the PSF is explicitly generated in chip coordinates, not sky coordinates, so modeling that
part of the PSF in sky coordinates means that the WCS (including tree rings) is being applied where it should not be,
leading to a signature of the WCS in the size residuals.
However, we also note that \cite{Magnier18} have identified variations in the charge diffusion size itself
associated with the same doping variations that cause the astrometric tree-ring effect
in the Pan-STARRS1 CCDs.
This effect may be present in DECam CCDs as well, which could be
contributing to the residuals seen in \fig{fig:fov2}.

When we develop the full optical plus atmospheric PSF model (cf. \sect{sec:future:composite}), 
we plan to include the possibility of having
a Gaussian component in CCD coordinates applied at the end to better model this effect.

\subsection{Residuals by magnitude}
\label{sec:mag}
 
\begin{figure}
\includegraphics[width=0.4 \textwidth]{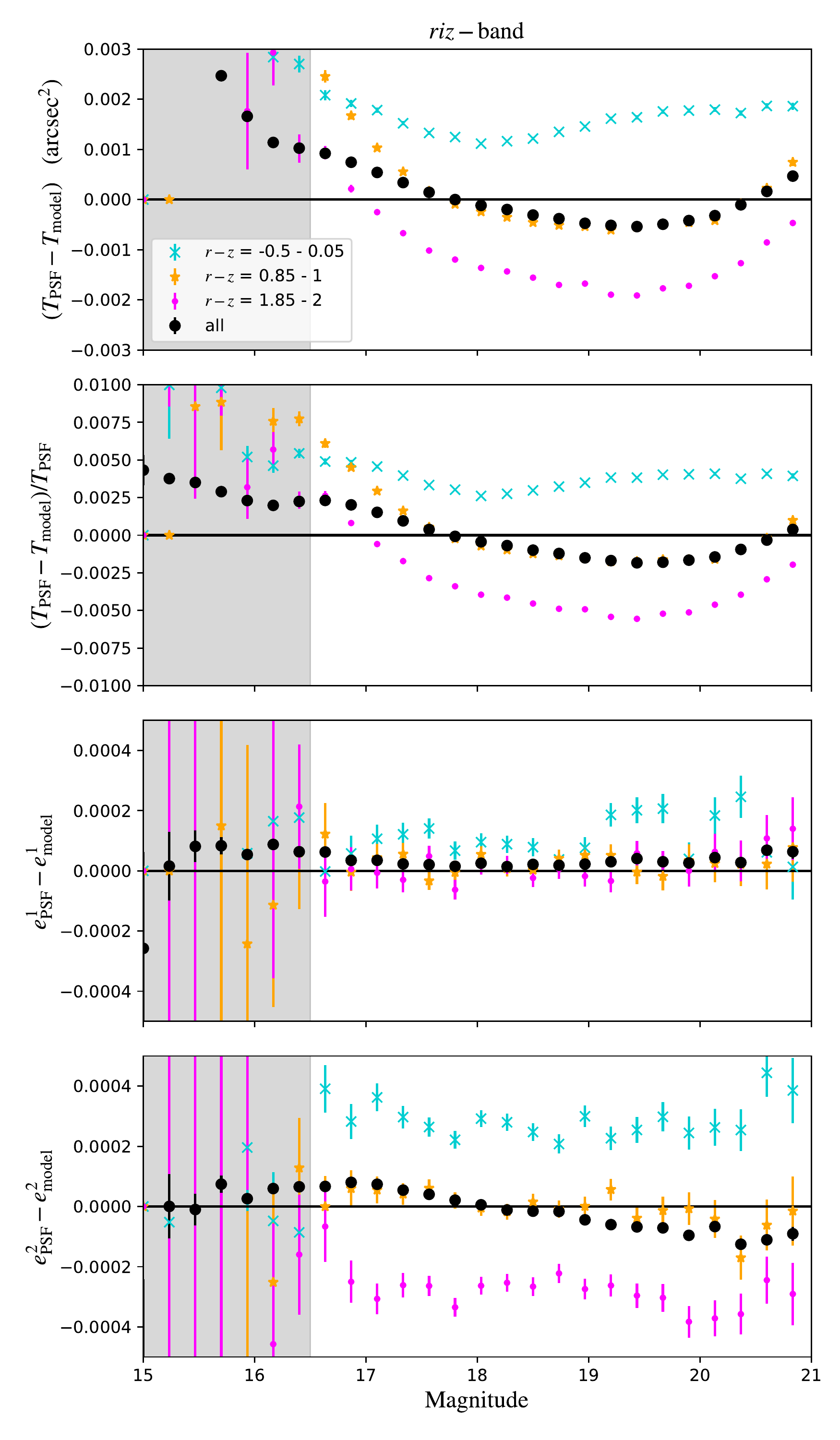}
\caption{The PSF residual size (top), fractional size (middle) and shape (bottom two) of stars as a function of their magnitude. 
In each case the ``PSF'' subscript refers to the true PSF, as estimated from direct measurements of images
of reserve stars, and ``model'' refers to the \piff\ estimate of the PSF at those locations.
The black points show the average values for all reserve stars in the $r$, $i$, and $z$ bands, which are
the bands used for cosmic shear measurements in the DES Y3 WL analysis.  Cyan, orange, and pink points
show the average values in thin slices of $r-z$ color, ranging from blue to red.
To reduce the impact of the brighter-fatter effect bright stars are excluded from our PSF models; 
the exact cut-off varies among CCD exposures but the shaded grey region shows a typical example.
}
\label{fig:mag}
\end{figure}

As discussed above in \sect{sec:bf}, an important feature of CCD data is the 
``brighter-fatter effect'' \citep{Antilogus14, Guyonnet15},
where the charge of electrons accumulated in high-flux pixels repels some of the other electrons
arriving later in the exposure, causing
bright objects to appear somewhat larger than they would otherwise have appeared.
Without any correction, this leads to an obvious trend of PSF size with magnitude.
The effect is also anisotropic, which leads to a strong trend in the $e_1$ shape residuals
as well \citep*{y1-shearcat}.
As we have discussed, the Y3 data reduction process included a correction for this effect,
which we can test by looking at PSF size and shape residuals as a function of magnitude.

The black points in \fig{fig:mag} show size residuals (upper panel), fractional size residuals (second panel), 
and $e_1$ and $e_2$ shape residuals (lower two panels) of the $riz$ reserve stars
as a function of their magnitude. 
The size residual is well below the level measured in Y1, and shows very little trend with magnitude,
remaining below 0.5\% over the entire range.
The shape residuals also show no significant trend with magnitude.  This implies that the brighter-fatter
correction we applied, along with the bright star cut, is sufficient to remove the impact of this effect
on the PSF solutions.

\subsection{Trends with color}
\label{sec:color}

There are several physical effects that are expected to cause the PSF to be wavelength-dependent
\citep{Plazas12, Meyers15}.
The PSF size from Kolmogorov seeing is expected to vary as
$\lambda^{-0.2}$ \citep[p. 92]{Hardy98}
\edit{or even steeper when taking into account the so-called outer scale \citep{Xin18}}.
Differential chromatic diffraction (DCR) causes the PSF to spread along the direction towards
zenith, affecting bluer stars more than redder stars.
Diffraction and other optical aberrations generally scale proportionally with $\lambda$.  There are a few refractive
elements, which have a non-trivial wavelength dependence.  And the conversion depth of photons
in the silicon increases with wavelength, which affects the PSF size,
\edit{due to the fast beam leading to a shallow depth of focus}.

We do not explicitly include any of these effects in the PSF modeling for the Y3 analysis
(although see \sect{sec:future:chromatic} for discussion of how we plan to include color dependence
in the future).
We thus expect the size residuals to be a function of the color of the stars.

The colored points in \fig{fig:mag} show the size and shape residuals for three thin slices
in $r-z$ color.  Cyan shows the bluest stars, pink the reddest, and orange in between.
The size residuals show a very clear trend with color.  The blue stars are larger than the average
model, and the red stars are smaller.  This implies
\edit{that the atmosphere (which causes red stars to appear smaller)}
is probably dominating over other effects (which mostly cause the size to
increase with $\lambda$). 

This can thus cause a bias in the inferred shapes of galaxies if the mean galaxy color is
significantly different from the mean color of the stars used to constrain the PSF.
See \citet*{y3-shapecatalog} for further discussion of the impact of this on the sample of galaxies
used for the Y3 weak lensing shear catalog.

The $e_1$ residuals (third panel of \fig{fig:mag}) show very little color dependence, but
the $e_2$ residuals (fourth panel of \fig{fig:mag}) do show a significant color dependence.
The redder stars have a
smaller than average $e_2$ shape, and the bluer stars have a larger than average $e_2$.

\begin{figure}
\includegraphics[width=0.5 \textwidth]{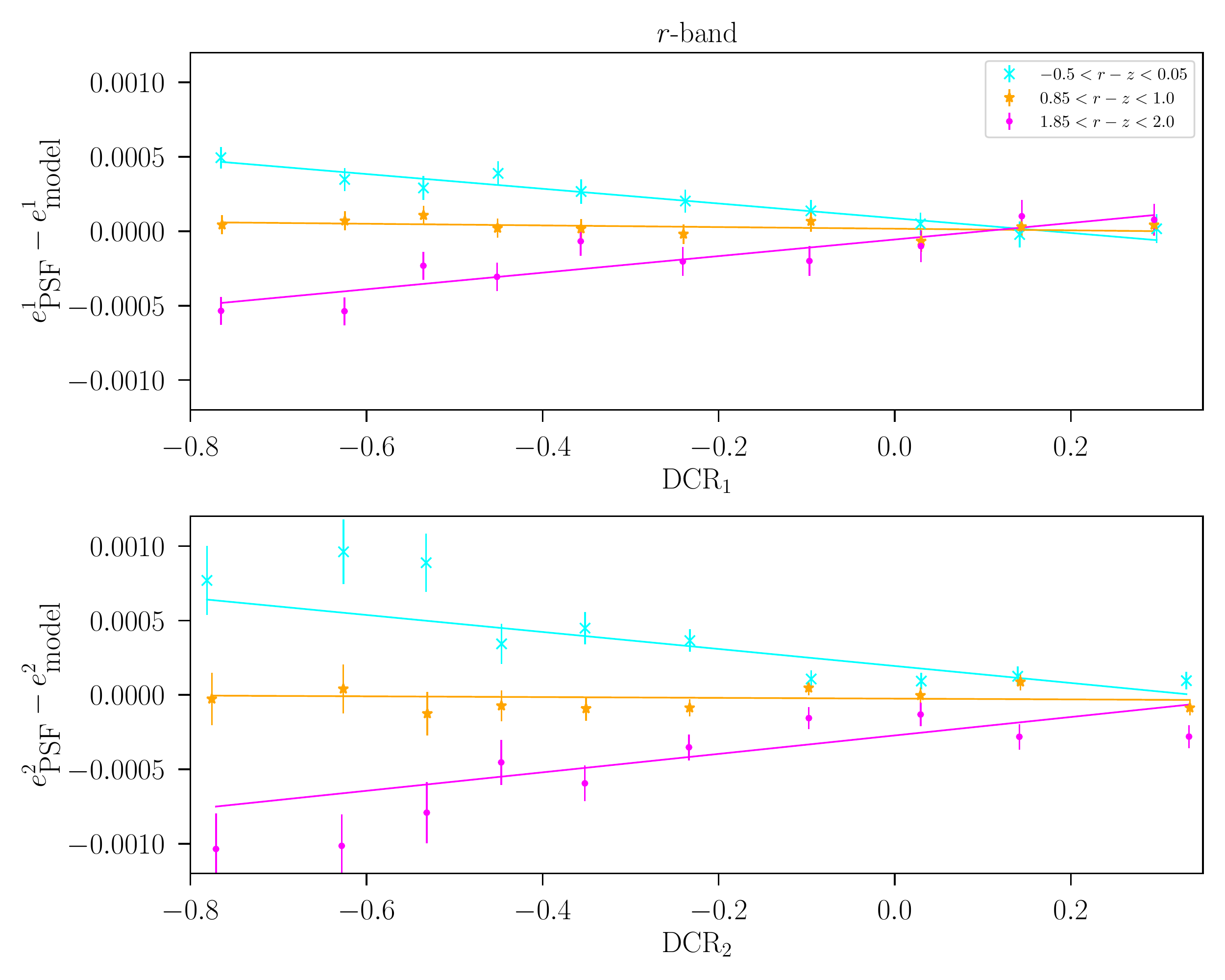}
\caption{
The PSF shape residuals in $r$-band
as a function of the mean direction and magnitude expected for differential chromatic
refraction (DCR).
The three sets of points correspond to the same three thin slices of $r-z$ color as shown in \fig{fig:mag}.
}
\label{fig:dcr}
\end{figure}

\edit{
This trend seems to be primarily due to DCR.  We have calculated the expected direction and magnitude
of the DCR effect across the DES Y3 footprint, quantified by two numbers:
\begin{align}
\mathrm{DCR}_1 &= \tan^2(z) \cos(2 q) \\
\mathrm{DCR}_2 & = \tan^2(z) \sin(2 q)
\end{align}
where $z$ is the zenith angle and $q$ is the parallactic angle.  We calculate the weighted mean
of these numbers for each location on the sky, 
based on the observations that contributed to the co-add images \citep{y3-gold}, 
using \textsc{HealSparse}\footnote{
\url{https://healsparse.readthedocs.io/en/latest/}} for efficient access.
\fig{fig:dcr} shows the mean shape residuals binned by these DCR quantities for just the
$r$-band observations where the DCR effect is strongest.  The three sets of points correspond
to the same $r-z$ colors as in \fig{fig:mag}.
The DES observing history (mostly
the particular history of the hour angle of each observation) happens
to have favored negative values of both quantities.  The binning scheme in \fig{fig:dcr} is
such that each point includes equal numbers of stars.  The mean residuals for the red and blue
color slices are close to
zero when the DCR quantities are near zero, and they are large when the DCR quantities
are large (in absolute value). 
The points do not perfectly follow the linear fits, but it is
clear that DCR is the dominant effect driving the difference between the residuals for 
red stars and blue stars.
The corresponding plots for $i$-band (not shown) are similar, but with smaller amplitude.
The ones for $z$-band show almost no effect at all.  These are what would be expected, since
the magnitude of the DCR effect decreases quickly with increasing wavelength.
}

\begin{figure*}
\includegraphics[width=\columnwidth]{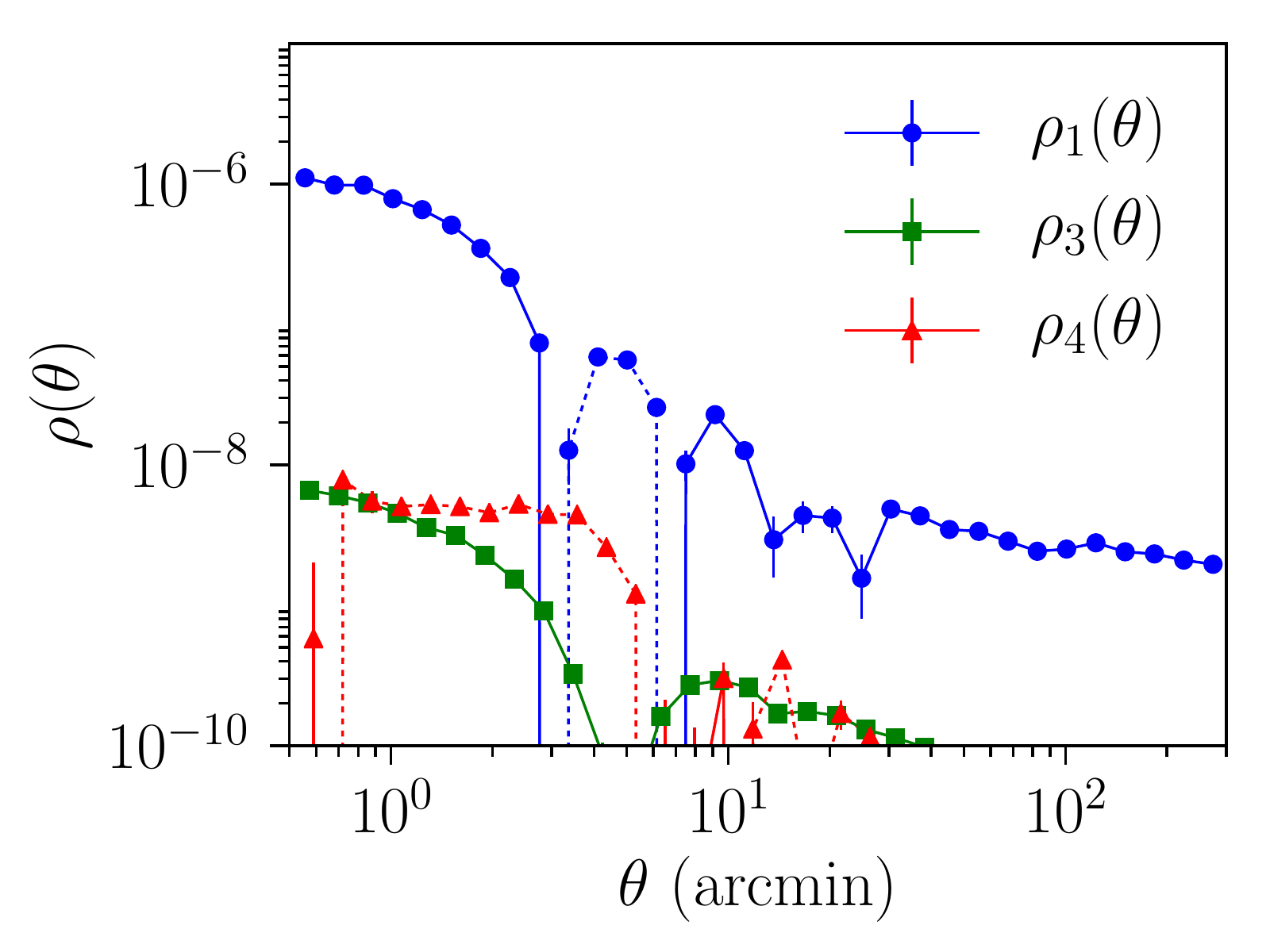}
\includegraphics[width=\columnwidth]{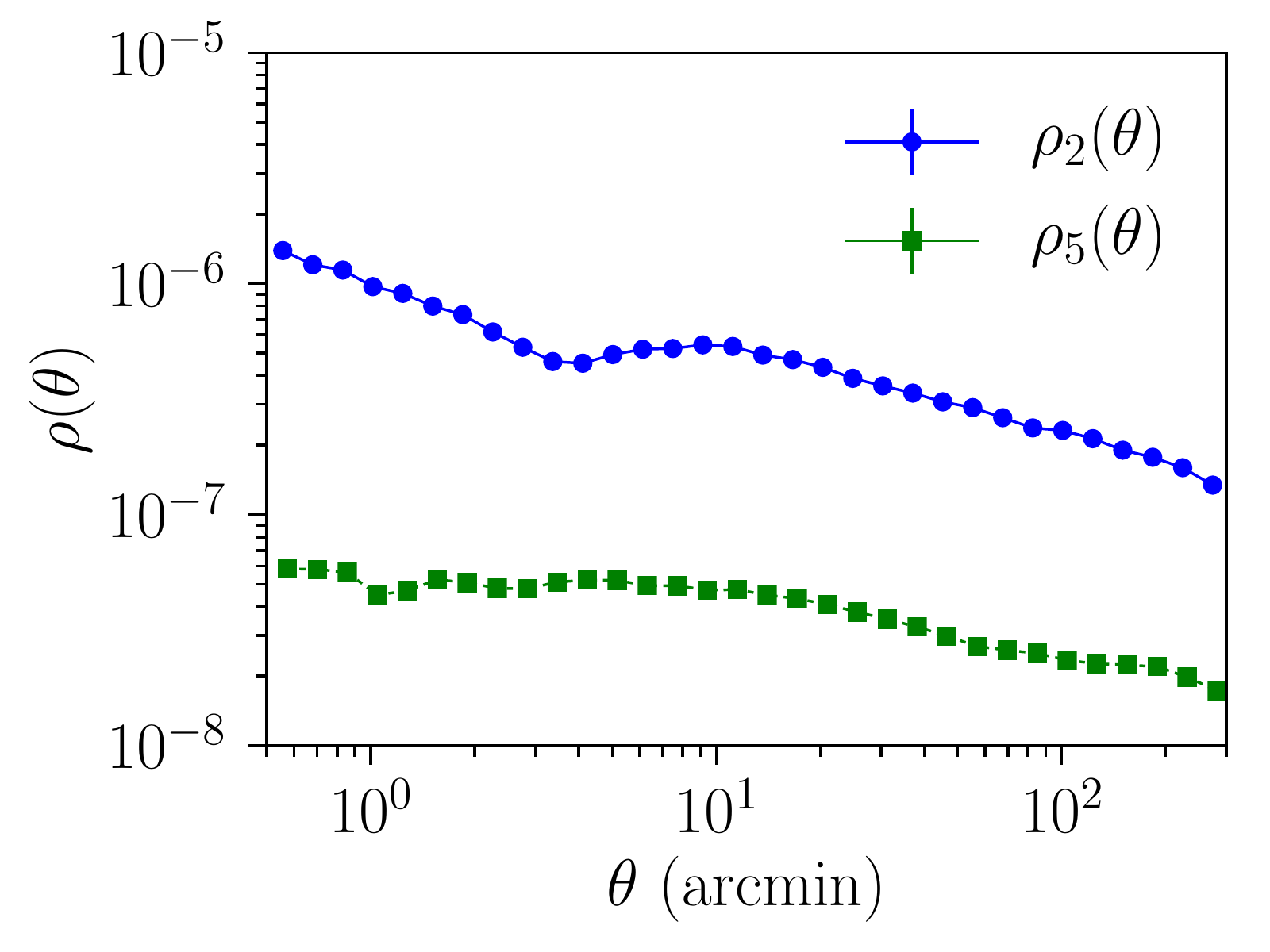}
\caption{
The $\rho$ statistics for the PSF residuals 
(\eqnc{eq:rho1}{eq:rho5}) in the $riz$ bands.
Negative values are shown in absolute value as dotted lines.
\label{fig:rho}
}
\end{figure*}

\begin{figure}
\includegraphics[width=0.5\columnwidth]{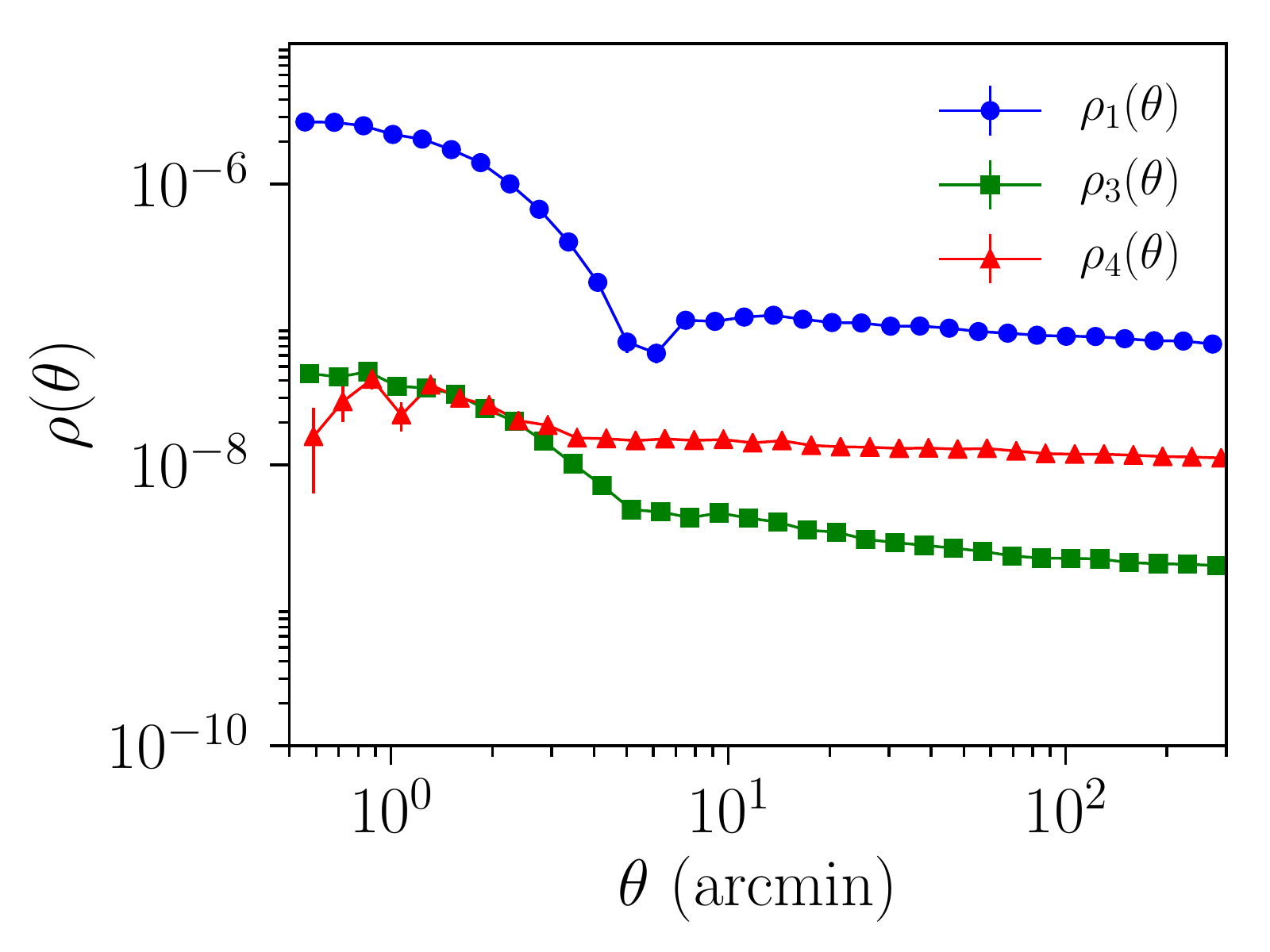}
\includegraphics[width=0.5\columnwidth]{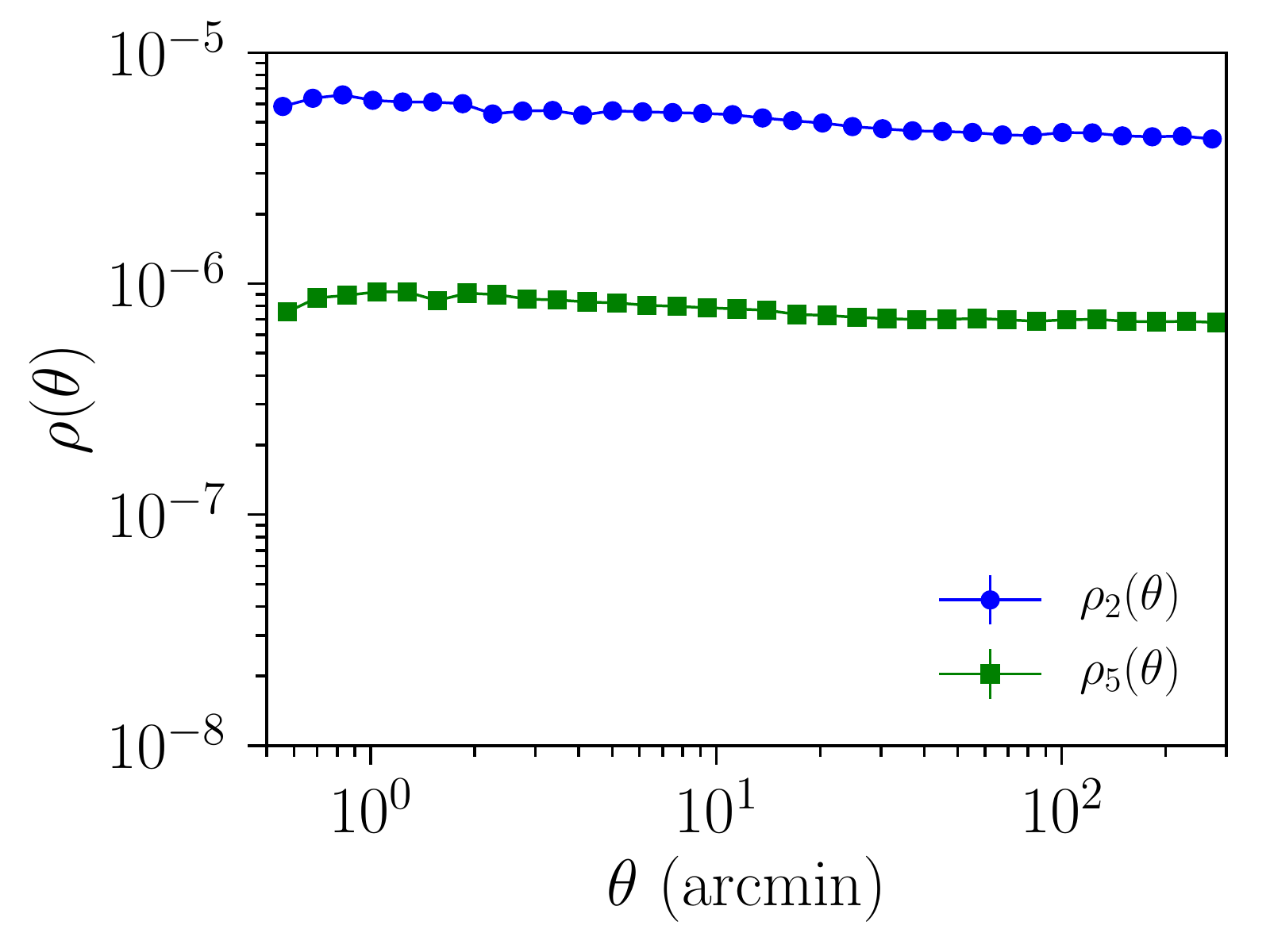}
\caption{
The $\rho$ statistics for the PSF residuals in the $g$ band.
These statistics are much worse that than the ones for the
$riz$ bands shown in \fig{fig:rho}.
Possible reasons for this are given in the text.  These results contributed to
the decision not to use the $g$ band for DES Y3 WL analysis.
\label{fig:rhog}
}
\end{figure}

\subsection{Rho statistics}
\label{sec:rho}

\citet{Rowe10} introduced the original two rho statistics for quantifying the spatial correlations of errors
in PSF models, $\rho_1(\theta)$ and $\rho_2(\theta)$.  \citet{sv-shearcat} developed three more related statistics that appear
at the same order in their potential impact on weak lensing two-point correlation functions:
\begin{linenomath*}
\begin{align}
\label{eq:rho1}
\rho_1(\theta) &\equiv \left \langle \delta \epsf^*(\bfx) \delta \epsf(\bfxpt) \right \rangle \\
\label{eq:rho2}
\rho_2(\theta) &\equiv \left \langle \epsf^*(\bfx) \delta \epsf(\bfxpt) \right \rangle \\
\label{eq:rho3}
\rho_3(\theta) &\equiv \left \langle \left(\epsf^* \frac{\delta\Tpsf}{\Tpsf}\right)\!(\bfx)
   \left(\epsf \frac{\delta\Tpsf}{\Tpsf}\right)\!(\bfxpt) \right \rangle \\
\label{eq:rho4}
\rho_4(\theta) &\equiv \left \langle \delta\epsf^*(\bfx)
   \left(\epsf \frac{\delta\Tpsf}{\Tpsf}\right)\!(\bfxpt) \right \rangle \\
\label{eq:rho5}
\rho_5(\theta) &\equiv \left \langle \epsf^*(\bfx)
   \left(\epsf \frac{\delta\Tpsf}{\Tpsf}\right)\!(\bfxpt) \right \rangle
\end{align}
\end{linenomath*}
where $\epsf$ is the observed ellipticity of the PSF stars,
$\Tpsf$ is the observed size of the PSF stars,
$\delta \epsf$ is the difference between the measured ellipticity of the observed stars and
the ellipticity of the \piff\ models at the same locations,
and $\delta\Tpsf$ is the difference in the sizes of the observed stars and the \piff\ models.

These statistics, if non-zero, imply some systematic errors in the weak lensing shear correlation
function, $\xi_+$.  There are corresponding statistics for $\xi_-$ contamination, but we find
these to be negligible, so we focus our attention henceforth on these five rho statistics.

The five rho statistics for the $riz$ bands are shown in \fig{fig:rho}.  
The left panel shows $\rho_1$, $\rho_3$, and $\rho_4$, which
represent direct systematic errors in $\xi_+$ with a leading
coefficient of order unity \citep{sv-shearcat}.  
The right panel shows $\rho_2$ and $\rho_5$,
whose impact on $\xi_+$ is mediated by a coefficient, $\alpha$,
describing the amount of ``PSF leakage'' that occurs during the shear measurement
process.

All of the rho statistics are small enough that they are not expected to cause significant
systematic errors in the cosmic shear measurements.  In particular, $\rho_1$ is about
a factor of $10$ smaller at large scales and a factor of $4$ smaller at small scales
that what was achieved for the Y1 PSF solution \citep*[Figure 9]{y1-shearcat}.
The impact of this on $\xi_+$ had been one of the largest systematic uncertainties
in the Y1 analysis.  

At large scales, $\rho_2$ is not much smaller than we found for the Y1 analysis, 
but its impact is also
expected to be small because our shear measurement method
has very little PSF leakage.
The estimated value of the leakage parameter, $\alpha$, for the DES Y3 WL
analysis is of order $10^{-2}$.  See \cite*{y3-shapecatalog} for details.

The rho statistics for each of $r$, $i$, and $z$ bands separately (not shown)
are somewhat noisier, but none of them show any particular problems.
On the other hand, the $g$-band rho statistics, shown in \fig{fig:rhog}
are an order of magnitude or more
larger for all five statistics, particularly at large scales.  
This could be due to the increased impact of
DCR in the $g$ band, or the non-stellar contamination in $g$ band discussed in
\sect{sec:purity}, or possibly other factors.  Regardless of the
cause, these rho statistics were considered unacceptably high, and
we decided to exclude the $g$-band data from our Y3 weak
lensing analysis.

\section{Planned Future Improvements}
\label{sec:future}

We have several plans for how the PSF solution could be further improved
for the final DES year six analysis (Y6) as well as for other surveys
with similar or smaller statistical errors.
These features are all currently being developed for release
in a future version of \piff.

\subsection{DECam Optical Model}
\label{sec:future:optical}

We have been developing an optical model of the Blanco telecope and DECam focal plane
to directly predict the effects of the optics on the PSF pattern.  A preliminary version of this
model was described in \cite{Davis16}.  The model is based on measurements of images
of stars in out-of-focus exposures, which produce ``donut'' images.  These reference wavefront
images allow for very
precise estimation of the Zernike decomposition of the optical wavefront as a function
of position in the focal plane, with only a small number of degrees of freedom  for how these
Zernike patterns change from one exposure to another.

This model has now been implemented in \piff, with either 10 or 11\footnote{The amplitude
of the spherical aberration may be either fixed or fitted.}
free parameters that are fit from a given in-focus image.  Most of the Zernike coefficients are
obtained
from the reference wavefront images and are not fitted using the in-focus image.  
We generally include coefficients up to
Zernike order 38 in the reference information, but even higher order can be accommodated
if desired.

Three of the degrees of freedom are currently
the average size and shape of the atmospheric component of the PSF, which are not part of the optical model.  
This makes the
current implementation of the optical model very accurate on average, but not particularly
accurate in the variation across any single exposure, since the atmospheric component
is the dominant contributor to this variation.  This leads us to our next planned improvement.

\subsection{Composite PSF}
\label{sec:future:composite}

We have started to develop a composite PSF class in \piff\ with 
multiple components convolved together
to produce the complete PSF solution.  Each component will be able to be separately
fit, using different models and different interpolation schemes as appropriate for each.

This composite class is primarily intended to allow for the convolution of an optical model and an
atmospheric model.  These two components of the PSF are constrained in very different
ways.  The optical model, as described above, has very few free parameters to be fit for
a given in-focus exposure, but its model of the surface brightness is quite complex
including very high order Zernike aberrations.

The atmospheric component, on the other hand, is fairly simple in its description at the 
location of a single star, being well approximated by an elliptical Kolmogorov
or von Karman profile.
However, the variation of the PSF parameters across the field of view is quite complicated.
Gaussian process interpolation is generally found to work well, which involves a
large matrix inversion to solve for the relevant parameters.

We also expect that a third Gaussian component, modeling charge diffusion inside the
CCD, will also be important to include, since this component interacts differently with the
WCS (especially the tree rings) than the other two components (cf. \sect{sec:fov}).
This component would probably only require a single fitted parameter for the entire
focal plane, being the average scale size of the Gaussian diffusion.

Development of a composite model that combines all three components in a single
iteration cycle is still ongoing.  We are hopeful that we will have this working in time to
use it for DES Y6 analysis.

\subsection{Using Gaia stars as input}
\label{sec:future:gaia}

We have seen in \sect{sec:purity} that the Gaia stellar catalog is somewhat shallower
than the input stellar catalog we used for the DES Y3 WL analysis.  
Using this
as our input stellar catalog would result in losing about half of the potential stars
for constraining the PSF.

We are nonetheless
considering switching to using that as our input catalog for Y6.  The most obvious
advantage to this is that it would avoid any concerns about galaxies leaking into
the stellar sample (especially in $g$ band where this was a particular problem).
The Gaia catalog constitutes an extremely pure stellar sample.
Also, the stars that would be included are mostly high signal-to-noise;
the stars we would be losing have less useful information about the PSF.

However, the bigger advantage is that the Gaia stars include extremely precise
astrometric information.  We could therefore switch to the ``fixed-star'' mode described
in \sect{sec:scheme}, where we would fix the true positions of the stars at the
Gaia positions (taking into account parallax and proper motion to find the position
at the time of each DES exposure) and allow the observed PSF profile to
include a small centroid offset.

These centroid offsets are expected from the effects of atmospheric seeing.
The atmospheric component of the PSF is integrated over a finite exposure time,
and the net integrated pattern includes a small shift in the observed centroid.
Furthermore, we expect these centroid offsets to be well fit by the same kind
of Gaussian process interpolation that is effective for interpolating the size and shape
parameters.

We thus plan to investigate including this centroid offset as part of the
atmospheric component of the PSF to see if it can improve the overall
astrometric modeling of the galaxies.

\subsection{Including chromatic dependence}
\label{sec:future:chromatic}

We saw in \sect{sec:mag} that the size and shape residuals vary with the
color of the stars.  This is completely expected, since several of the physical
effects that make up the PSF pattern are wavelength-dependent.

For the Y6 PSF solution, 
we plan to include a single color parameter (e.g. $g{-}r$) in the fit to account for
\edit{the wavelength dependence on the PSF to first order}.
Then when using the
PSF model for galaxies, the color of the galaxy would need to be provided to 
produce an estimate of the correct PSF for that specific galaxy.
We are hopeful that including this color dependence will improve the $g$-band
PSF solution sufficiently to allow it to be used for weak lensing in the Y6 analysis.

We designed the \piff\ API to allow for this kind of color term to be included in the
interpolation scheme, anticipating this use case.  However, we have not yet tried using this
functionality on real data, so there will likely be some development required to get
it working on DES Y6 data.

\section{Summary}
\label{sec:conclusion}

We have presented a new software package for PSF estimation of astronomical
images, called \piff, which was developed primarily for the DES Y3 WL
analysis.  The \piff\ PSF models were tested on the Y3 data and show significantly smaller residuals
than had been seen in the Y1 data.
Most notably, the $\rho_1$ statistic is more than an order of magnitude smaller at large scales than
what was found for the Y1 PSF model.  This had been one of the most significant sources of
systematic uncertainty in the Y1 cosmic shear analysis.

Development of \piff\ is still very active.  We have described in \sect{sec:future} several potential improvements we
hope to include in the DES Y6 analysis.  In addition, as part of the 
Dark Energy Science Collaboration (DESC) for the Vera Rubin Observatory Legacy Survey of Space and
Time (LSST), we have been testing \piff\ on simulated LSST images.  Results of these tests
will be reported separately, but we fully expect that \piff\ will prove to be useful for
LSST and other surveys in addition to DES.  Indeed, the design of \piff\ is very general, so we
expect that it will be useful to many other current and future surveys who need accurate
PSF modeling based on observations of stars.

\section*{Data Availability}

The \piff\ software is publicly available\footnote{\url{https://github.com/rmjarvis/Piff}} under an
open source license.  
This paper describes version 1.0 of the software, although most of the tests were made
with PSF models produced by version 0.2.4.  
Installation instructions are found on the website, and we welcome
feature requests and bug reports from users.
The code we used to run \piff\ on DES images,
produce the PSF catalogs, and create many of the plots is also publicly 
available\footnote{\url{https://github.com/rmjarvis/DESWL/tree/master/psf} and 
\url{https://github.com/des-science/y3psfpaper}}.
Catalogs of the PSF measurements on the reserve stars will be made available as part of
the DES Y3 coordinated release\footnote{\url{https://des.ncsa.illinois.edu/releases}}.

\section*{Acknowledgements}

This paper has gone through internal review by the DES collaboration.  We thank
the internal reviewers for their very helpful suggestions for improving the paper.
\edit{We also thank Robert Lupton for a very careful and helpful review as referee
for this paper.}

M. Jarvis and G. Bernstein are partially supported by the US Department of Energy
grant DE-SC0007901 and funds from the University of Pennsylvania.

This document was prepared by the DES collaboration using the resources of the
Fermi National Accelerator Laboratory (Fermilab), a U.S. Department of Energy, Office of Science,
HEP User Facility. Fermilab is managed by Fermi Research Alliance, LLC (FRA),
acting under Contract No. DE-AC02-07CH11359.

This work was based in part on observations at Cerro Tololo Inter-American Observatory,
National Optical Astronomy Observatory, which is operated by the Association of
Universities for Research in Astronomy (AURA) under a cooperative agreement with the National
Science Foundation.

Funding for the DES Projects has been provided by the U.S. Department of Energy, the U.S. National Science Foundation, the Ministry of Science and Education of Spain,
the Science and Technology Facilities Council of the United Kingdom, the Higher Education Funding Council for England, the National Center for Supercomputing
Applications at the University of Illinois at Urbana-Champaign, the Kavli Institute of Cosmological Physics at the University of Chicago,
the Center for Cosmology and Astro-Particle Physics at the Ohio State University,
the Mitchell Institute for Fundamental Physics and Astronomy at Texas A\&M University, Financiadora de Estudos e Projetos,
Funda{\c c}{\~a}o Carlos Chagas Filho de Amparo {\`a} Pesquisa do Estado do Rio de Janeiro, Conselho Nacional de Desenvolvimento Cient{\'i}fico e Tecnol{\'o}gico and
the Minist{\'e}rio da Ci{\^e}ncia, Tecnologia e Inova{\c c}{\~a}o, the Deutsche Forschungsgemeinschaft and the Collaborating Institutions in the Dark Energy Survey.

The Collaborating Institutions are Argonne National Laboratory, the University of California at Santa Cruz, the University of Cambridge, Centro de Investigaciones Energ{\'e}ticas,
Medioambientales y Tecnol{\'o}gicas-Madrid, the University of Chicago, University College London, the DES-Brazil Consortium, the University of Edinburgh,
the Eidgen{\"o}ssische Technische Hochschule (ETH) Z{\"u}rich,
Fermi National Accelerator Laboratory, the University of Illinois at Urbana-Champaign, the Institut de Ci{\`e}ncies de l'Espai (IEEC/CSIC),
the Institut de F{\'i}sica d'Altes Energies, Lawrence Berkeley National Laboratory, the Ludwig-Maximilians Universit{\"a}t M{\"u}nchen and the associated Excellence Cluster Universe,
the University of Michigan, the National Optical Astronomy Observatory, the University of Nottingham, The Ohio State University, the University of Pennsylvania, the University of Portsmouth,
SLAC National Accelerator Laboratory, Stanford University, the University of Sussex, Texas A\&M University, and the OzDES Membership Consortium.

\bibliography{literature,des,des_y1kp,des_y3kp}

\appendix


\section{Constrained Centroid Method for PixelGrid}
\label{constrainedcentroid}

The version of \piff\ used for the DES Y3 WL analysis (0.2.4) included a different
mechanism for constraining the centroid of the \code{PixelGrid} model
to be $(0,0)$ than we now employ (version 0.3.0 and later).  For transparency reasons, we describe
the older method here.

As described in \sect{sec:pixelgrid}, the \code{PixelGrid} PSF model is
\begin{linenomath*}
\begin{align}
I(u,v) &= \sum_{k=1}^{N_\mathrm{pix}} p_k K(u{-}u_k) K(v{-}v_k)
\end{align}
\end{linenomath*}
where $K(x) {=} L_n(x)$ is the Lanczos interpolation kernel (\eqn{eq:lanczos}).
The coefficients $p_k$ for a given star
can be constrained by minimizing
\begin{linenomath*}
\begin{align}
\chi^2 &= \sum_{\alpha} 
\frac{\left(d_\alpha - f A_\mathrm{pix} I(u_\alpha{-}u_c,v_\alpha{-}v_c) \right)^2}
{\sigma_\alpha^2 + d_\alpha}
\label{eq:chisq_app}
\end{align}
\end{linenomath*}
where the sum on $\alpha$ is over the observed data pixels and the star has some flux $f$ and
centroid $(u_c,v_c)$. 
Minimizing this leads to a design matrix for the coefficients $\{p_k\}$
\begin{linenomath*}
\begin{align}
\mathbf{A}_0 \mathbf{p} &= \mathbf{b}_0
\end{align}
\end{linenomath*}

If we want to force the solution to have zero centroid and unit flux, we can impose
three additional constraint equations on the solution:
\begin{linenomath*}
\begin{align}
\sum_k \begin{pmatrix} 1 \\ u_i \\ v_i \end{pmatrix} p_k &= \begin{pmatrix} 1 \\ 0 \\ 0 \end{pmatrix} \\
\mathbf{A}_1 \mathbf{p} &= \mathbf{b}_1
\label{eq:constraint}
\end{align}
\end{linenomath*}
To allow \eqn{eq:constraint} to hold, our fit must have three fewer degrees of freedom.
Three of the coefficients in $\mathbf{p}$ need to be determined directly from the rest using \eqn{eq:constraint}.
We denote the three coefficients to be excluded from the fit as $\mathbf{p}_c$, the constrained
coefficients, which in practice are taken to be the central pixel and two of its neighbors.  
The other coefficients, denoted $\mathbf{p}_f$, are the fitted coefficients.  

Without loss of generality, we can arrange the constrained coefficients to have index $0,1,2$, and
the fitted coefficients to have index $3\, ..\, N_\mathrm{pix}{-}1$.  Then \eqn{eq:constraint} becomes
\begin{linenomath*}
\begin{align}
\mathbf{A}_1[:,0:3] \mathbf{p}_c + \mathbf{A}_1[:,3:N_\mathrm{pix}] \mathbf{p}_f &= \mathbf{b}_1
\end{align}
\end{linenomath*}
which can be directly solved for $\mathbf{p}_c$ given a solution $\mathbf{p}_f$.

According to Bayes Theorem, the pixel data $d_\alpha$ for a given star
constrain $\mathbf{p}_f$, $f$, $u_c$ and $v_c$ according to the likelihood:
\begin{linenomath*}
\begin{align}
\mathcal{L}(\mathbf{p}_f, f, u_c, v_c) &\propto e^{-\chi^2/2} P_f(f) P_u(u_c) P_v(v_c)
\label{bayes}
\end{align}
\end{linenomath*}
where $P_f$, $P_u$, and $P_v$ are priors on the flux and centroid parameters and
$\chi^2$ is from \eqn{eq:chisq_app}.
We used Gaussian priors in all cases.  For the flux, we assumed a Gaussian width of 0.5 times
the current flux estimate.  For the centroid parameters, we assumed a width of 0.5 pixels.

Note that $\chi^2$ is quadratic in $\mathbf{p}_f$, and therefore the minimum is linear in $\mathbf{p}_f$.
The only non-linear terms arise from the flux and centroid parameters.  However, these should generally
be small adjustments during the iterative solution, so we can linearize this equation in $\delta f, \delta u_c, \delta v_c$.
This then allows for these three parameters to be marginalized analytically producing a linear
design equation
\begin{linenomath*}
\begin{align}
\mathbf{A} \delta \mathbf{p}_f &= \mathbf{b}
\end{align}
\end{linenomath*}
for each iteration of the solution.

\section{Configuration Used for DES Y3 PSF Model}
\label{config}

As mentioned in the main text, we used \piff\ version 0.2.4 for the DES Y3 PSF solution.
The input configuration file was the following:

\begin{lstlisting}[language=yaml]
modules:
    - galsim_extra

input:
    image_file_name: # Set on command line
    image_hdu: 0
    badpix_hdu: 1
    weight_hdu: 2

    cat_file_name: # Set on command line
    cat_hdu: 1
    x_col: x
    y_col: y
    sky_col: sky
    ra: TELRA
    dec: TELDEC

    gain: 1.0
    max_snr: 100
    min_snr: 20
    stamp_size: 25

    wcs:
        type: Pixmappy
        file_name: # Set on command line
        exp: # Set on command line
        ccdnum: # Set on command line

output:
    file_name: # Set on command line

psf:
    model:
        type: PixelGrid
        scale: 0.30
        size: 17
    interp:
        type: BasisPolynomial
        order: 3
    outliers:
        type: Chisq
        nsigma: 5.5
        max_remove: 0.01
\end{lstlisting}

Of course, some values needed to be set differently for each exposure and CCD, so these were specified
on the command line when we ran the \code{piffify} executable for each image.  The other parameters
shown here were the same for all exposures.

\section{Author Affiliations}
\label{sec:affiliations}
$^{1}$ Department of Physics and Astronomy, University of Pennsylvania, Philadelphia, PA 19104, USA\\
$^{2}$ Kavli Institute for Particle Astrophysics \& Cosmology, P. O. Box 2450, Stanford University, Stanford, CA 94305, USA\\
$^{3}$ LPNHE, CNRS/IN2P3, Sorbonne Universit\'e, Laboratoire de Physique Nucl\'eaire et de Hautes \'Energies, F-75005, Paris, France\\
$^{4}$ Physics Department, 2320 Chamberlin Hall, University of Wisconsin-Madison, 1150 University Avenue Madison, WI  53706-1390\\
$^{5}$ Jodrell Bank Center for Astrophysics, School of Physics and Astronomy, University of Manchester, Oxford Road, Manchester, M13 9PL, UK\\
$^{6}$ Institut de F\'{\i}sica d'Altes Energies (IFAE), The Barcelona Institute of Science and Technology, Campus UAB, 08193 Bellaterra (Barcelona) Spain\\
$^{7}$ SLAC National Accelerator Laboratory, Menlo Park, CA 94025, USA\\
$^{8}$ Department of Astronomy and Astrophysics, University of Chicago, Chicago, IL 60637, USA\\
$^{9}$ Kavli Institute for Cosmological Physics, University of Chicago, Chicago, IL 60637, USA\\
$^{10}$ Department of Physics, Duke University Durham, NC 27708, USA\\
$^{11}$ Center for Cosmology and Astro-Particle Physics, The Ohio State University, Columbus, OH 43210, USA\\
$^{12}$ Department of Physics, IIT Hyderabad, Kandi, Telangana 502285, India\\
$^{13}$ Fermi National Accelerator Laboratory, P. O. Box 500, Batavia, IL 60510, USA\\
$^{14}$ Department of Physics, Stanford University, 382 Via Pueblo Mall, Stanford, CA 94305, USA\\
$^{15}$ Department of Astronomy, University of Illinois at Urbana-Champaign, 1002 W. Green Street, Urbana, IL 61801, USA\\
$^{16}$ National Center for Supercomputing Applications, 1205 West Clark St., Urbana, IL 61801, USA\\
$^{17}$ Department of Physics, The Ohio State University, Columbus, OH 43210, USA\\
$^{18}$ Lawrence Livermore National Laboratory, 7000 East Avenue, Livermore, CA 94550, USA\\
$^{19}$ Instituto de F\'isica Gleb Wataghin, Universidade Estadual de Campinas, 13083-859, Campinas, SP, Brazil\\
$^{20}$ Laborat\'orio Interinstitucional de e-Astronomia - LIneA, Rua Gal. Jos\'e Cristino 77, Rio de Janeiro, RJ - 20921-400, Brazil\\
$^{21}$ Department of Astrophysical Sciences, Princeton University, Peyton Hall, Princeton, NJ 08544, USA\\
$^{22}$ Brookhaven National Laboratory, Bldg 510, Upton, NY 11973, USA\\
$^{23}$ Madison West High School, 30 Ash St, Madison, WI 53726, USA\\
$^{24}$ Institute for Astronomy, University of Edinburgh, Edinburgh EH9 3HJ, UK\\
$^{25}$ Cerro Tololo Inter-American Observatory, NSF's National Optical-Infrared Astronomy Research Laboratory, Casilla 603, La Serena, Chile\\
$^{26}$ Departamento de F\'isica Matem\'atica, Instituto de F\'isica, Universidade de S\~ao Paulo, CP 66318, S\~ao Paulo, SP, 05314-970, Brazil\\
$^{27}$ Instituto de Fisica Teorica UAM/CSIC, Universidad Autonoma de Madrid, 28049 Madrid, Spain\\
$^{28}$ Department of Physics and Astronomy, Pevensey Building, University of Sussex, Brighton, BN1 9QH, UK\\
$^{29}$ Department of Physics \& Astronomy, University College London, Gower Street, London, WC1E 6BT, UK\\
$^{30}$ Instituto de Astrofisica de Canarias, E-38205 La Laguna, Tenerife, Spain\\
$^{31}$ Universidad de La Laguna, Dpto. Astrofísica, E-38206 La Laguna, Tenerife, Spain\\
$^{32}$ INAF-Osservatorio Astronomico di Trieste, via G. B. Tiepolo 11, I-34143 Trieste, Italy\\
$^{33}$ Institute for Fundamental Physics of the Universe, Via Beirut 2, 34014 Trieste, Italy\\
$^{34}$ Observat\'orio Nacional, Rua Gal. Jos\'e Cristino 77, Rio de Janeiro, RJ - 20921-400, Brazil\\
$^{35}$ Centro de Investigaciones Energ\'eticas, Medioambientales y Tecnol\'ogicas (CIEMAT), Madrid, Spain\\
$^{36}$ Santa Cruz Institute for Particle Physics, Santa Cruz, CA 95064, USA\\
$^{37}$ Institut d'Estudis Espacials de Catalunya (IEEC), 08034 Barcelona, Spain\\
$^{38}$ Institute of Space Sciences (ICE, CSIC),  Campus UAB, Carrer de Can Magrans, s/n,  08193 Barcelona, Spain\\
$^{39}$ Department of Astronomy, University of Michigan, Ann Arbor, MI 48109, USA\\
$^{40}$ Department of Physics, University of Michigan, Ann Arbor, MI 48109, USA\\
$^{41}$ School of Mathematics and Physics, University of Queensland,  Brisbane, QLD 4072, Australia\\
$^{42}$ Center for Astrophysics $\vert$ Harvard \& Smithsonian, 60 Garden Street, Cambridge, MA 02138, USA\\
$^{43}$ Australian Astronomical Optics, Macquarie University, North Ryde, NSW 2113, Australia\\
$^{44}$ Lowell Observatory, 1400 Mars Hill Rd, Flagstaff, AZ 86001, USA\\
$^{45}$ George P. and Cynthia Woods Mitchell Institute for Fundamental Physics and Astronomy, and Department of Physics and Astronomy, Texas A\&M University, College Station, TX 77843,  USA\\
$^{46}$ Instituci\'o Catalana de Recerca i Estudis Avan\c{c}ats, E-08010 Barcelona, Spain\\
$^{47}$ Institute of Astronomy, University of Cambridge, Madingley Road, Cambridge CB3 0HA, UK\\
$^{48}$ School of Physics and Astronomy, University of Southampton,  Southampton, SO17 1BJ, UK\\
$^{49}$ Computer Science and Mathematics Division, Oak Ridge National Laboratory, Oak Ridge, TN 37831\\
$^{50}$ Max Planck Institute for Extraterrestrial Physics, Giessenbachstrasse, 85748 Garching, Germany\\
$^{51}$ Universit\"ats-Sternwarte, Fakult\"at f\"ur Physik, Ludwig-Maximilians Universit\"at M\"unchen, Scheinerstr. 1, 81679 M\"unchen, Germany\\

\label{lastpage}

\end{document}